# Power-Bandwidth Tradeoff in Dense Multi-Antenna Relay Networks

Özgür Oyman, *Member, IEEE*, and Arogyaswami J. Paulraj, *Fellow, IEEE*

***Abstract*— We consider a dense fading multi-user network with multiple active multi-antenna source-destination pair terminals communicating simultaneously through a large common set of $K$ multi-antenna relay terminals in the full spatial multiplexing mode. We use Shannon-theoretic tools to analyze the tradeoff between energy efficiency and spectral efficiency (known as the power-bandwidth tradeoff) in meaningful asymptotic regimes of signal-to-noise ratio (SNR) and network size. We design linear distributed multi-antenna relay beamforming (LDMRB) schemes that exploit the spatial signature of multi-user interference and characterize their power-bandwidth tradeoff under a system-wide power constraint on source and relay transmissions. The impact of multiple users, multiple relays and multiple antennas on the key performance measures of the high and low SNR regimes is investigated in order to shed new light on the possible reduction in power and bandwidth requirements through the usage of such practical relay cooperation techniques. Our results indicate that point-to-point coded multi-user networks supported by distributed relay beamforming techniques yield enhanced energy efficiency and spectral efficiency, and with appropriate signaling and sufficient antenna degrees of freedom, can achieve asymptotically optimal power-bandwidth tradeoff with the best possible (i.e., as in the cutset bound) energy scaling of $K^{-1}$ and the best possible spectral efficiency slope at any SNR for large number of relay terminals. Furthermore, our results help to identify the role of interference cancellation capability at the relay terminals on realizing the optimal power-bandwidth tradeoff; and show how relaying schemes that do not attempt to mitigate multi-user interference, despite their optimal capacity scaling performance, could yield a poor power-bandwidth tradeoff.**

## I. Introduction

The design of large-scale wireless distributed (adhoc) networks poses a set of new challenges to information theory, communication theory and network theory. Such networks are characterized by the large size of the network both in terms of the number of nodes (i.e., *dense*) and in terms of the geographical area the network covers. Furthermore, each terminal could be severely constrained by its computational and transmission/receiving power and/or scarcity of bandwidth resources. These constraints require an understanding of the performance limits of such networks *jointly in terms of energy efficiency and spectral efficiency*. This paper applies tools from information theory and statistics to evaluate the performance limits of dense wireless networks focusing on the tradeoff between energy efficiency and spectral efficiency; which is also known as the *power-bandwidth tradeoff*.

**Relation to Previous Work.** While the power-bandwidth tradeoff characterizations of various point-to-point and multi-user communication settings can be found in the literature, previous work addressing the fundamental limits over large adhoc wireless networks has focused either only on the energy efficiency performance [1] or only on the spectral efficiency performance [2]-[6]. The analytical tools to study the power-bandwidth tradeoff in the power-limited regime have been previously developed in the context of point-to-point single-user communications [7]-[8], and were extended to multi-user (point-to-multipoint and multipoint-to-point) settings [9]-[12], as well as to adhoc wireless networking examples of single-relay channels [13]-[14], single-antenna relay networks [15] and linear multi-hop networks [16]-[17]. In the bandwidth-limited regime, the necessary tools to perform the power-bandwidth tradeoff analysis were developed by [9] in the context of code-division multiple access (CDMA) systems and were later used by [18] and [19] to characterize fundamental limits in multi-antenna channels [20]-[26] over point-to-point and broadcast communication, respectively, and by [17] to study the end-to-end performance of multi-hop routing techniques over linear non-ergodic fading wireless networks.

**Contributions.** This paper characterizes the power-bandwidth tradeoff over dense fixed-area wireless adhoc networks in the limit of large number of terminals, in the special case of a multi-antenna relay network (MRN) model. In particular, we design low-complexity linear distributed multi-antenna relay beamforming (LDMRB) schemes that take advantage of local channel state information (CSI) to convey simultaneously multiple users' signals to their intended destinations and quantify enhancements in energy efficiency and spectral efficiency achievable from such practical relay cooperation schemes. We remark that some of the results to be presented here have appeared before in [27]-[29]. Our key findings can be summarized as follows:

Manuscript received October 10, 2005; revised May 27, 2006; accepted October 6, 2006. The associate editor coordinating the review of this paper and approving it for publication was Q. Zhang. This paper was presented in part at the $39^{\rm th}$ Conference on Information Sciences and Systems, Baltimore, MD, March 2005, the $43^{\rm rd}$ Allerton Conference on Communications, Control, and Computing, Monticello, IL, Sep. 2005, and the $1^{\rm st}$ IEEE Workshop on Wireless Mesh Networks, Santa Clara, CA, Sep. 2005.

This work was supported by the NSF grant DMS-0354674-001 and ONR grant N00014-02-0088.

Ö. Oyman was with the Information Systems Laboratory, Stanford University, Stanford, CA 94305, U.S.A. He is now with the Corporate Technology Group, Intel Corporation, Santa Clara, CA 95054, U.S.A. (email: ozgur.oyman@intel.com)

A. J. Paulraj is with the Information Systems Laboratory, Stanford University, Stanford, CA 94305, U.S.A. (email: apaulraj@stanford.edu)

- LDMRB is *asymptotically optimal* for any SNR in point-to-point coded multi-user MRNs. In particular, we prove that with bursty signaling, much better energy scaling ($K^{-1}$ rather than $K^{-1/2}$) is achievable with LDMRB compared to previous work in [1] and we verify the optimality of the $K^{-1}$ energy scaling by analyzing the cutset upper bound [30] on the multi-user MRN spectral efficiency in the limit of large number of relay terminals. Furthermore, we show that LDMRB simultaneously achieves the best possible spectral efficiency slope (i.e., as upper bounded by the cutset theorem) at any SNR.
- Interference cancellation capability at the relay terminals plays a key role in achieving the optimal power-bandwidth tradeoff. Our results demonstrate how LDMRB schemes that do not attempt to mitigate multi-user interference, despite their optimal capacity scaling performance, could be energy inefficient; and yield a poor power-bandwidth tradeoff in the high SNR regime due to the interference-limited nature of the multi-user MRN.

The rest of this paper is organized as follows. Section II describes the multi-user MRN model and the power-bandwidth tradeoff problem formulation. In Section III, we derive an upper-limit on the achievable power-bandwidth tradeoff using the cut-set theorem [30]. We analyze the performance of MRN for various LDMRB schemes in Section IV. Finally, we present our numerical results in Section V and conclude in Section VI.

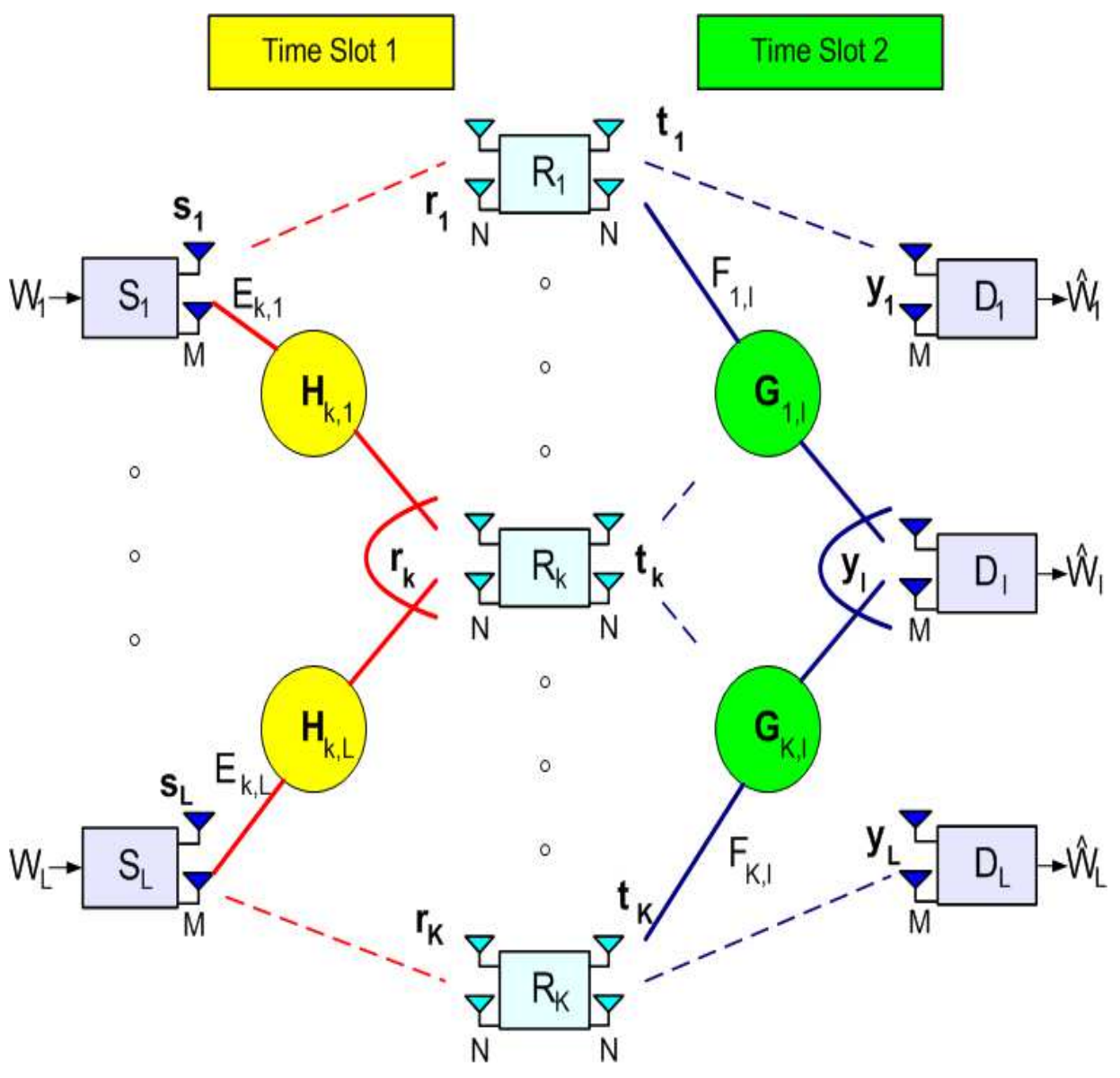

Fig. 1. Multi-user MRN source-relay and relay-destination channel models.

## II. Network Model and Definitions

**General Assumptions.** We assume that the MRN consists of $K+2L$ terminals, with $L$ active source-destination pairs and $K$ relay terminals located randomly and independently in a domain of fixed area. We denote the $l$-th source terminal by $\mathcal{S}_l$, the $l$-th destination terminal by $\mathcal{D}_l$, where $l=1,...,L$, and the $k$-th relay terminal by $\mathcal{R}_k,\ k=1,2,...,K$. The source and destination terminals $\{\mathcal{S}_l\}$ and $\{\mathcal{D}_l\}$ are equipped with $M$ antennas each, while each of the relay terminals $\mathcal{R}_k$ employs $N$ transmit/receive antennas. We assume that there is a "dead zone" of non-zero radius around $\{\mathcal{S}_l\}$ and $\{\mathcal{D}_l\}$ [3], which is free of relay terminals and that no direct link exists between the source-destination pairs. The source terminal $\mathcal{S}_l$ is only interested in sending data to the destination terminal $\mathcal{D}_l$ by employing point-to-point coding techniques (without any cooperation across source-destination pairs) and the communication of all $L$ source-destination pairs is supported through the same set of $K$ relay terminals. As terminals can often not transmit and receive at the same time, we consider time-division based (half duplex) relaying schemes for which transmissions take place in two hops over two separate time slots. In the first time slot, the relay terminals receive the signals transmitted from the source terminals. After processing the received signals, the relay terminals simultaneously transmit their data to the destination terminals during the second time slot.

**Channel and Signal Model.** Throughout the paper, frequency-flat fading over the bandwidth of interest and perfectly synchronized transmission/reception between the terminals is assumed. In case of frequency-selective fading, the channel can be decomposed into parallel non-interacting subchannels each experiencing frequency-flat fading and having the same Shannon capacity as the overall channel. The channel model is depicted in Fig. 1. The discrete-time complex baseband input-output relation for the $\mathcal{S}_l \rightarrow \mathcal{R}_k$ link [1] over the first time-slot is given by

$$\mathbf{r}_k = \sum_{l=1}^{L} \sqrt{E_{k,l}}\, \mathbf{H}_{k,l}\mathbf{s}_l + \mathbf{n}_k, \quad k=1,2,...,K,$$

where $\mathbf{r}_k \in \mathbb{C}^N$ is the received vector signal at $\mathcal{R}_k$, $E_{k,l} \in \mathbb{R}$ is the scalar energy normalization factor to account for path loss and shadowing in the $\mathcal{S}_l \rightarrow \mathcal{R}_k$ link, $\mathbf{H}_{k,l} \in \mathbb{C}^{N\times M}$ is the corresponding channel matrix independent across source and relay terminals (i.e., independent across $k$ and $l$) and consisting of i.i.d. $\mathcal{CN}(0,1)$ entries, $\mathbf{s}_l \in \mathbb{C}^M$ is the spatio-temporally i.i.d. (i.e., assuming full spatial multiplexing [31] for all multi-antenna transmissions; which implies that $M$ independent spatial streams are sent simultaneously by each $M$-antenna source terminal) zero-mean circularly symmetric complex Gaussian transmit signal vector for $\mathcal{S}_l$ satisfying $\mathbb{E}\left[\mathbf{s}_l\mathbf{s}_l^H\right] = (P_{\mathcal{S}_l}/M)\,\mathbf{I}_M$ (i.e. $P_{\mathcal{S}_l} = \mathbb{E}\left[\|\mathbf{s}_l\|^2\right]$ is the average transmit power for source terminal $\mathcal{S}_l$), and $\mathbf{n}_k \in \mathbb{C}^N$ is the spatio-temporally white zero-mean circularly symmetric complex Gaussian noise vector at $\mathcal{R}_k$, independent across $k$, with single-sided noise power spectral density $N_0$.

As part of LDMRB, each relay terminal $\mathcal{R}_k$ *linearly* processes its received vector signal $\mathbf{r}_k$ to produce the vector signal $\mathbf{t}_k \in \mathbb{C}^N$ (i.e., $\exists\, \mathbf{A}_k \in \mathbb{C}^{N\times N}$ such that $\mathbf{t}_k = \mathbf{A}_k\mathbf{r}_k,\ \forall k$), which is then transmitted to the destination terminals over the second time slot.[2] The destination terminal $\mathcal{D}_l$ receives the

[1] $\mathcal{A} \rightarrow \mathcal{B}$ signifies communication from terminal $\mathcal{A}$ to terminal $\mathcal{B}$.

[2] In the presence of linear beamforming at the relay terminals, the source-destination links $\mathcal{S}_l \rightarrow \mathcal{D}_l,\ l=1,...,L$ can be viewed as a composite *interference channel* [30] where the properties of the resulting conditional channel distribution function $p(\{\mathbf{y}_{l,m}\}\,|\,\{\mathbf{s}_{l,m}\})$ rely upon the choice of the LDMRB matrices $\{\mathbf{A}_k\}_{k=1}^K$.

signal vector $\mathbf{y}_l \in \mathbb{C}^M$ expressed as

$$\mathbf{y}_l = \sum_{k=1}^{K} \sqrt{F_{k,l}}\, \mathbf{G}_{k,l}\mathbf{t}_k + \mathbf{z}_l, \quad l = 1, ..., L,$$

where $F_{k,l} \in \mathbb{R}$ is the scalar energy normalization factor to account for path loss and shadowing in the $\mathcal{R}_k \to \mathcal{D}_l$ link, $\mathbf{G}_{k,l} \in \mathbb{C}^{M\times N}$ is the corresponding channel matrix with i.i.d. $\mathcal{CN}(0,1)$ entries, independent across $k$ and $l$, and $\mathbf{z}_l \in \mathbb{C}^M$ is the spatio-temporally white circularly symmetric complex Gaussian noise vector at $\mathcal{D}_l$ with single-sided noise power spectral density $N_0$. The transmit signal vector $\mathbf{t}_k$ satisfies the average power constraint $\mathbb{E}\left[\|\mathbf{t}_k\|^2\right] \leq P_{\mathcal{R}_k}$ ($P_{\mathcal{R}_k}$ is the average transmit power for relay terminal $\mathcal{R}_k$). [3]

As already mentioned above, throughout the paper, the path-loss and shadowing statistics are captured by $\{E_{k,l}\}$ (for the first hop) and $\{F_{k,l}\}$ (for the second hop). We assume that these parameters are random, i.i.d., strictly positive (due to the fact that the domain of interest has a fixed area, i.e. dense network), bounded above (due to the dead zone requirement), and remain constant over the entire time period of interest. Additionally, we assume an ergodic block fading channel model such that the channel matrices $\{\mathbf{H}_{k,l}\}$ and $\{\mathbf{G}_{k,l}\}$ remain constant over the entire duration of a time slot and change in an independent fashion across time slots. Finally, we assume that there is no CSI at the source terminals $\{\mathcal{S}_l\}$, each relay terminal $\mathcal{R}_k$ has perfect knowledge of its local forward and backward channels, $\{F_{k,l}, \mathbf{G}_{k,l}\}_{l=1}^{L}$ and $\{E_{k,l}, \mathbf{H}_{k,l}\}_{l=1}^{L}$, respectively, and the destination terminals $\{\mathcal{D}_l\}$ have perfect knowledge of all channel variables.[4]

**Channel Coding Framework.** For any block length $Q$, a $(\{2^{QR_{l,m}} : l = 1, ..., L, m = 1, ..., M\}, Q)$ code $\mathcal{C}_Q$ is defined such that $R_{l,m}$ is the rate of communication over the $m$-th spatial stream of the $l$-th source-destination pair. In this setting, all multi-antenna transmissions employ full spatial multiplexing and horizontal encoding/decoding [31]. The source codebook for the multi-user MRN (of size $\sum_{l=1}^{L}\sum_{m=1}^{M} 2^{QR_{l,m}}$ codewords) is determined by the encoding functions $\{\phi_{l,m}\}$ that map each message $w_{l,m} \in \mathcal{W}_{l,m} = \{1, ..., 2^{QR_{l,m}}\}$ of $\mathcal{S}_l$ to a transmit codeword $\mathbf{s}_{l,m} = [\, s_{l,m,1}, \ldots, s_{l,m,Q}\,] \in \mathbb{C}^Q$, where $s_{l,m,q} \in \mathbb{C}$ is the transmitted symbol from antenna $m$ of $\mathcal{S}_l$ at time $q = 1, ..., Q$ (corresponding to the $m$-th spatial stream of $\mathcal{S}_l$). Under the two-hop relaying protocol, $Q$ symbols are transmitted over each hop for each of the $LM$ spatial streams. For the reception of the $m$-th spatial stream of source-destination pair $l$, destination terminal $\mathcal{D}_l$ employs a decoding function $\psi_{l,m}$ to perform the mapping $\mathbb{C}^Q \to \hat{w}_{l,m} \in \mathcal{W}_{l,m}$ based on its received signal $\mathbf{y}_{l,m} = [\, y_{l,m,1}, \ldots, y_{l,m,Q}\,]$, where $y_{l,m,q} \in \mathbb{C}$ is the received symbol at antenna $m$ of $\mathcal{D}_l$ at time $q+1$, i.e., due to communication over two hops, symbols transmitted by the source terminals at time $q$ are received by the destination terminals at time $q+1$. The error probability for the $m$-th spatial stream of the $l$-th source-destination pair is given by $\epsilon_{l,m} = \mathbb{P}(\psi_{l,m}(\mathbf{y}_{l,m}) \neq w_{l,m})$. The $LM$-tuple of rates $\{R_{l,m}\}$ is achievable if there exists a sequence of $(\{2^{QR_{l,m}}\}, Q)$ codes $\{\mathcal{C}_Q : Q = 1, 2, ...\}$ with vanishing $\epsilon_{l,m}$, $\forall l$, $\forall m$.

**Power-Bandwidth Tradeoff Measures.** We assume that the network is supplied with fixed finite total power $P$ over unconstrained bandwidth $B$. We define the network signal-to-noise ratio (SNR) for the $\mathcal{S}_l \to \mathcal{D}_l$, $l = 1, ..., L$ links as

$$\mathsf{SNR}_{\,\mathrm{network}} \doteq \frac{P}{N_0 B} = \frac{\sum_{l=1}^{L} P_{\mathcal{S}_l} + \sum_{k=1}^{K} P_{\mathcal{R}_k}}{2N_0 B},$$

where the factor of $1/2$ comes from the half duplex nature of source and relay transmissions. Note that our definition of network SNR captures power consumption at the relay as well as source terminals ensuring a fair performance comparison between distributed relaying and direct transmissions. To simplify notation, from now on we refer to $\mathsf{SNR}_{\mathrm{network}}$ as $\mathsf{SNR}$. Due to the statistical symmetry of their channel distributions, we allow for equal power allocation among the source and relay terminals and set $P_{\mathcal{S}_l} = P_{\mathcal{S}}, \forall l$ and $P_{\mathcal{R}_k} = P_{\mathcal{R}}, \forall k$.

The multi-user MRN with desired sum rate $R = \sum_{l=1}^{L}\sum_{m=1}^{M} R_{l,m}$ (the union of the set of achievable rate LM-tuples $\{R_{l,m}\}$ defines the capacity region) must respect the fundamental limit $R/B \leq \mathsf{C}\,(E_b/N_0)$, where $\mathsf{C}$ is the Shannon capacity (ergodic mutual information[5] ) (in bits/second/Hertz or b/s/Hz), which we will also refer as the spectral efficiency, and $E_b/N_0$ is the energy per information bit normalized by background noise spectral level, expressed as $E_b/N_0 = \mathsf{SNR}/C(\mathsf{SNR})$ [6]. There exists a tradeoff between the efficiency measures $E_b/N_0$ and $\mathsf{C}$ (known as the power-bandwidth tradeoff) in achieving a given target data rate. When $\mathsf{C} \ll 1$, the system operates in the *power-limited regime*; i.e., the bandwidth is large and the main concern is the limitation on power. Similarly, the case of $\mathsf{C} \gg 1$ corresponds to the *bandwidth-limited regime*. Tightly framing achievable performance, particular emphasis in our power-bandwidth tradeoff analysis is placed on the regions of low and high $E_b/N_0$.

*Low $E_b/N_0$ regime:* Defining $(E_b/N_0)_{\min}$ as the minimum system-wide $E_b/N_0$ required to convey any positive rate reliably, we have $(E_b/N_0)_{\min} = \min \mathsf{SNR}/C(\mathsf{SNR})$, over all $\mathsf{SNR} \geq 0$. In most of the scenarios we will consider, $E_b/N_0$ is minimized when $\mathsf{SNR}$ is low. This regime of operation is referred as the *wideband regime* in which the spectral efficiency $\mathsf{C}$ is near zero. We consider the first-order behavior of $\mathsf{C}$ as a function of $E_b/N_0$ in the wideband regime (i.e.,

[3]Under a general frequency-selective block-fading channel model, our assumptions imply that each relay terminal transmits the same power over all frequency subchannels and fading blocks (equal power allocation), while it should be noted that the availability of channel state information at the relays allows for designing relay power allocation strategies across frequency subchannels and fading blocks. However, as the results of [4] show, optimal power allocation at the relay terminals does not enhance the capacity scaling achieved by equal power allocation, and therefore our asymptotic results on the power-bandwidth tradeoff and the related scaling laws for the energy efficiency and spectral efficiency measures would remain the same under optimal power allocation at the relays.

[4]As we shall show in Section IV, the CSI knowledge at the destination terminals is not required for our results to hold in the asymptotic regime where the number of relays tends to infinity.

[5]We emphasize that due to the ergodicity assumption on the channel statistics, a Shannon capacity exists (this is obtained by averaging the total mutual information between the source and destination terminals over the statistics of the channel processes) for the multi-user MRN.

[6]The use of $C$ and $\mathsf{C}$ avoids assigning the same symbol to spectral efficiency functions of $\mathsf{SNR}$ and $E_b/N_0$.

$\mathsf{SNR} \to 0$) by analyzing the affine function (in decibels) [7]

$$10\log_{10}\frac{E_b}{N_0}(\mathsf{C}) = 10\log_{10}\frac{E_b}{N_0}_{\min} + \frac{\mathsf{C}}{S_0}10\log_{10}2 + o(\mathsf{C}),$$

where $S_0$ denotes the "wideband" slope of spectral efficiency in b/s/Hz/(3 dB) at the point $(E_b/N_0)_{\min}$,

$$S_0 = \lim_{\frac{E_b}{N_0}\downarrow\frac{E_b}{N_0}_{\min}} \frac{\mathsf{C}(\frac{E_b}{N_0})}{10\log_{10}\frac{E_b}{N_0} - 10\log_{10}\frac{E_b}{N_0}_{\min}} 10\log_{10}2.$$

It can be shown that [7]

$$\frac{E_b}{N_0}_{\min} = \lim_{\mathsf{SNR}\to 0}\frac{\ln 2}{\dot{C}(\mathsf{SNR})}, \quad \text{and} \quad S_0 = \lim_{\mathsf{SNR}\to 0}\frac{2\left[\dot{C}(\mathsf{SNR})\right]^2}{-\ddot{C}(\mathsf{SNR})}, \tag{1}$$

where $\dot{C}$ and $\ddot{C}$ denote the first and second order derivatives of $C(\mathsf{SNR})$ (evaluated in nats/s/Hz).

*High $E_b/N_0$ regime:* In the high SNR regime (i.e., $\mathsf{SNR} \to \infty$), the dependence between $E_b/N_0$ and $\mathsf{C}$ can be characterized as [9]

$$\begin{aligned} 10\log_{10}\frac{E_b}{N_0}(\mathsf{C}) &= \frac{\mathsf{C}}{S_\infty}10\log_{10}2 \\ &- 10\log_{10}(\mathsf{C}) + 10\log_{10}\frac{E_b}{N_0}_{\mathrm{imp}} + o(1), \end{aligned}$$

where $S_\infty$ denotes the "high SNR" slope of the spectral efficiency in b/s/Hz/(3 dB)

$$\begin{aligned} S_\infty &= \lim_{\frac{E_b}{N_0}\to\infty}\frac{\mathsf{C}(\frac{E_b}{N_0})}{10\log_{10}\frac{E_b}{N_0}}10\log_{10}2 \\ &= \lim_{\mathsf{SNR}\to\infty}\mathsf{SNR}\,\dot{C}(\mathsf{SNR}) \end{aligned} \tag{2}$$

and $(E_b/N_0)_{\mathrm{imp}}$ is the $E_b/N_0$ improvement factor with respect to a single-user single-antenna unfaded AWGN reference channel[8] and it is expressed as

$$\frac{E_b}{N_0}_{\mathrm{imp}} = \lim_{\mathsf{SNR}\to\infty}\left[\mathsf{SNR}\exp\left(-\frac{C(\mathsf{SNR})}{S_\infty}\right)\right]. \tag{3}$$

## III. Upper-Limit on MRN Power-Bandwidth Tradeoff

In this section, we derive an upper-limit on the achievable energy efficiency and spectral efficiency performance over the MRN, which will be key in the next section for establishing the asymptotic optimality of the MRN power-bandwidth tradeoff under LDMRB schemes. Based on the cut-set upper bound on network spectral efficiency, we now establish that the best possible energy scaling over a dense MRN is $K^{-1}$ at all SNRs and best possible spectral efficiency slopes are $S_0 = LM$ at low SNR and $S_\infty = LM/2$ at high SNR. It is clear that no capacity-suboptimal scheme (e.g., LDMRB) can yield a better power-bandwidth tradeoff.

**Theorem 1.** *In the limit of large $K$, $E_b/N_0$ can almost surely be lower bounded by*

$$\frac{E_b}{N_0}(\mathsf{C}) \geq \frac{2^{2\mathsf{C}(LM)^{-1}}-1}{2\mathsf{C}}\,\frac{LM}{KN\,\mathbb{E}\left[E_{k,l}\right]} + o\left(\frac{1}{K}\right). \tag{4}$$

[7] $u(x) = o(v(x)), x \to L$ stands for $\lim_{x\to L}\frac{u(x)}{v(x)} = 0$.

[8] For the AWGN channel; $C(\mathsf{SNR}) = \ln(1+\mathsf{SNR})$ resulting in $S_0 = 2$, $(E_b/N_0)_{\min} = \ln 2$, $S_\infty = 1$ and $(E_b/N_0)_{\mathrm{imp}} = 1$.

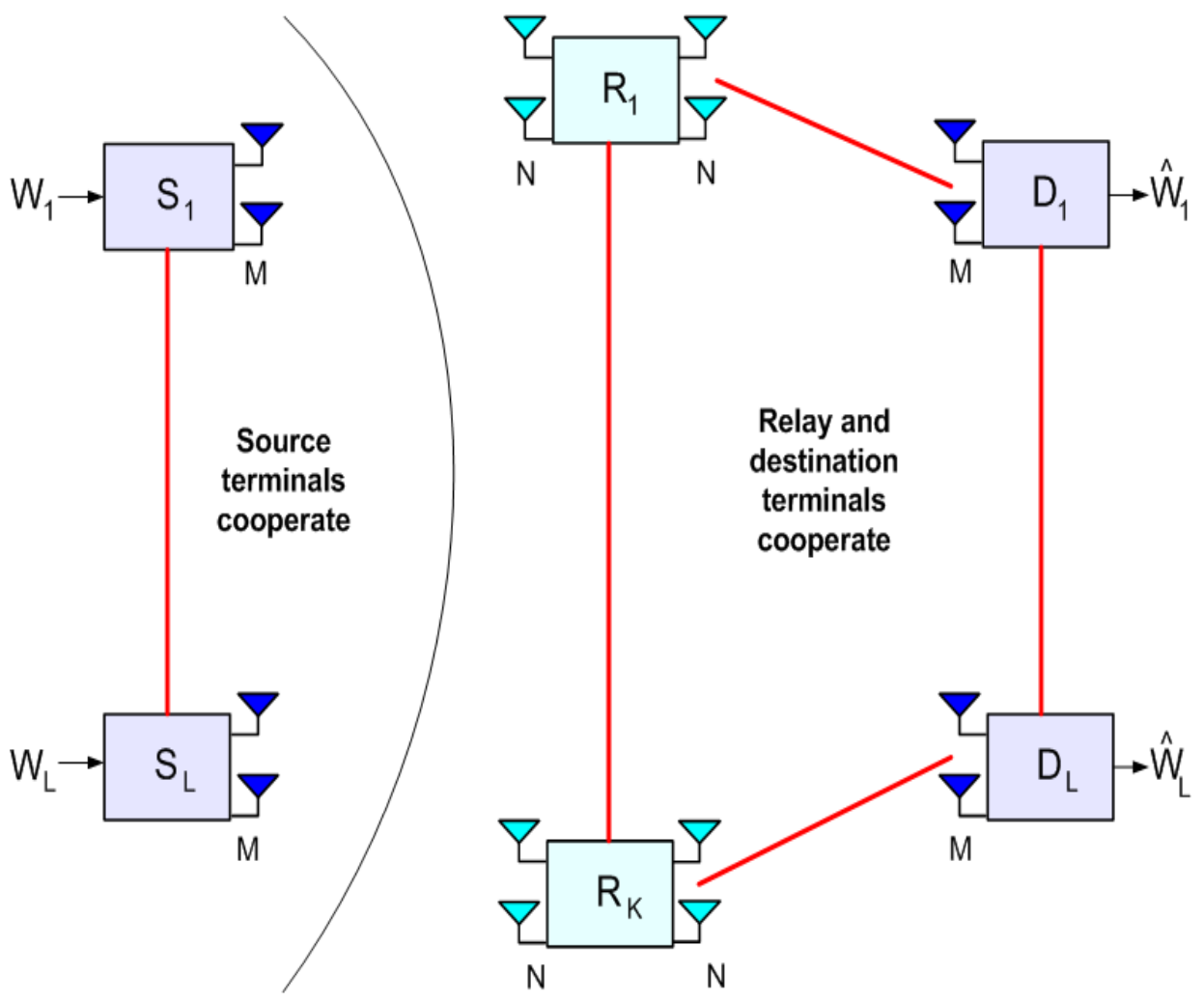

Fig. 2. Illustration of the broadcast cut over the MRN.

*1- Best-case power-bandwidth tradeoff at low $E_b/N_0$:*

$$\frac{E_b}{N_0}_{\min}^{\mathrm{best}} = \frac{\ln 2}{KN\,\mathbb{E}\left[E_{k,l}\right]} + o\left(\frac{1}{K}\right) \quad \textit{and} \quad S_0^{\mathrm{best}} = LM,$$

*2- Best-case power-bandwidth tradeoff at high $E_b/N_0$:*

$$\frac{E_b}{N_0}_{\mathrm{imp}}^{\mathrm{best}} = \frac{LM}{2KN\,\mathbb{E}\left[E_{k,l}\right]} + o\left(\frac{1}{K}\right) \quad \textit{and} \quad S_\infty^{\mathrm{best}} = \frac{LM}{2}.$$

**Proof:** Separating the source terminals $\{\mathcal{S}_l\}$ from the rest of the network using a broadcast cut (see Fig. 2), and applying the cut-set theorem (Theorem 14.10.1 of [30]), it follows that the spectral efficiency of the multi-user MRN can be upper bounded as

$$\mathsf{C} \leq \mathbb{E}_{\{\mathbf{H}_{k,l},\mathbf{G}_{k,l}\}}\left[\frac{1}{2}I(\{\mathbf{s}_l\}_{l=1}^{L};\{\mathbf{r}_k\}_{k=1}^{K},\{\mathbf{y}_l\}_{l=1}^{L}|\{\mathbf{t}_k\}_{k=1}^{K})\right],$$

where the factor $1/2$ results from the fact that data is transmitted over two time slots. Observing that in our network model $\{\mathbf{s}_l\} \to \{\mathbf{r}_k\} \to \{\mathbf{t}_k\} \to \{\mathbf{y}_l\}$ forms a Markov chain, applying the chain rule of mutual information [30] and using the fact that conditioning reduces entropy, we extend the upper bound to

$$\mathsf{C} \leq \mathbb{E}_{\{\mathbf{H}_{k,l}\}}\left[\frac{1}{2}I(\mathbf{s}_1,...,\mathbf{s}_L;\mathbf{r}_1,...,\mathbf{r}_K)\right].$$

Recalling that $\{\mathbf{s}_l\}$ are circularly symmetric complex Gaussian with $\mathbb{E}\left[\mathbf{s}_l\mathbf{s}_l^H\right] = (P_{\mathcal{S}}/M)\,\mathbf{I}_M$, we have

$$\mathsf{C} \leq \mathbb{E}_{\{\mathbf{H}_{k,l}\}}\left[\frac{1}{2}\log_2\left(\left|\mathbf{I}_{LM} + \frac{P_{\mathcal{S}}}{M\,N_0\,B}\mathbf{\Phi}\right|\right)\right], \tag{5}$$

where $\mathbf{\Phi} \in \mathbb{C}^{LM\times LM}$ is of the form

$$\mathbf{\Phi} = \begin{bmatrix} \mathbf{\Delta}_{1,1} & \cdots & \mathbf{\Delta}_{1,L} \\ \vdots & & \vdots \\ \mathbf{\Delta}_{L,1} & \cdots & \mathbf{\Delta}_{L,L} \end{bmatrix},$$

with matrices $\mathbf{\Delta}_{i,j} \in \mathbb{C}^{M\times M}$ given by

$$\mathbf{\Delta}_{i,j} = \sum_{k=1}^{K}\sqrt{E_{k,i}E_{k,j}}\mathbf{H}_{k,i}^H\mathbf{H}_{k,j}, \quad i = 1,...,L,\ j = 1,...,L$$

Now, applying Jensen's inequality to (5) it follows that

$$\mathsf{C} \le \frac{M}{2}\sum_{l=1}^{L}\log_2\left(1+\frac{P_{\mathcal{S}}\,N}{M\,N_0\,B}\sum_{k=1}^{K}E_{k,l}\right).$$

By our assumption that $\{E_{k,l}\}$ are bounded, it follows that $\{\mathrm{var}(E_{k,l})\}$ are also bounded $\forall k,\ \forall l$. Hence the Kolmogorov condition is satisfied and we can use Theorem 1.8.D of [32] to obtain [9]

$$\sum_{k=1}^{\infty}\frac{\mathrm{var}(E_{k,l})}{k^2}<\infty \quad\rightarrow\quad \sum_{k=1}^{K}\frac{E_{k,l}}{K}-\sum_{k=1}^{K}\frac{\mathbb{E}\left[E_{k,l}\right]}{K}\xrightarrow{\text{w.p.1}}0$$

resulting in (based on Theorem 1.7 of [32])

$$\mathsf{C} \le \frac{L\,M}{2}\log_2\left(1+\frac{P_{\mathcal{S}}KN\,\mathbb{E}\left[E_{k,l}\right]}{M\,N_0\,B}+o(K)\right) \tag{6}$$

as $K\rightarrow\infty$. Since our application of the cut-set theorem through the broadcast cut leads to perfect relay-destination (i.e. $\mathcal{R}_k\rightarrow\mathcal{D}_l$) links, relays do not consume any transmit power and hence we set $P_{\mathcal{R}}=0$ yielding $\mathsf{SNR}=\mathsf{C}\,\frac{E_b}{N_0}=L\,P_{\mathcal{S}}/(2N_0B)$. Substituting this relation into (6), we can show (4). Expressing the upper bound on $\mathsf{C}$ given in (6) in terms of $\mathsf{SNR}$ and applying (1)-(3), we complete the proof. □

## IV. MRN Power-Bandwidth Tradeoff with Practical LDMRB Techniques

In this section, we present practical (but suboptimal) LDMRB schemes such that each relay transmit vector $\mathbf{t}_k\in\mathbb{C}^N$ is a linear transformation of the corresponding received vector $\mathbf{r}_k\in\mathbb{C}^N$. These LDMRB schemes differ in the way they fight multi-stream interference (arising due to simultaneous transmission of multiple spatial streams from multiple source-destination pairs) and background Gaussian noise: *i) The matched filter (MF) algorithm* mitigates noise but ignores multi-stream interference. *ii) The zero-forcing (ZF) algorithm* cancels multi-stream interference completely (requiring $N\ge LM$), but amplifies noise. *iii) The linear minimum mean-square error (L-MMSE) algorithm* is the best tradeoff for interference and noise mitigation [31], [34]. The LDMRB schemes based on the ZF and L-MMSE algorithms have an interference mitigation advantage over the MF-based scheme in that they can exploit the differences in the spatial signatures of the interfering spatial streams to enhance the quality of the estimates on the desired spatial stream.

**LDMRB Schemes.** Each relay terminal exploits its knowledge of the local backward CSI $\{E_{k,l},\mathbf{H}_{k,l}\}_{l=1}^{L}$ to perform input linear-beamforming operations on its received signal vector to obtain estimates for each of the $LM$ transmitted spatial streams. Accordingly, terminal $\mathcal{R}_k$ correlates its received signal vector $\mathbf{r}_k$ with each of the beamforming (row) vectors $\mathbf{u}_{k,l,m}\in\mathbb{C}^N$ to yield $\hat{s}_{k,l,m}=\mathbf{u}_{k,l,m},\mathbf{r}_k$ such that

$$\begin{aligned}\hat{s}_{k,l,m} &= \sqrt{E_{k,l}}\,\mathbf{u}_{k,l,m}\mathbf{h}_{k,l,m}s_{l,m}\\ &+\sum_{(p,q)\neq(l,m)}\sqrt{E_{k,p}}\,\mathbf{u}_{k,l,m}\mathbf{h}_{k,p,q}\,s_{p,q}\;+\;\mathbf{u}_{k,l,m}\,\mathbf{n}_k,\end{aligned}$$

as the estimate for $s_{l,m}$, where $s_{p,q}$ denotes the transmitted signal from the $q$-th antenna of source $\mathcal{S}_p$, $p=1,2,...,L$, $q=1,2,...,M$, and $\mathbf{h}_{k,p,q}$ is the $q$-th column of $\mathbf{H}_{k,p}$. Following this operation, $\mathcal{R}_k$ sets the average energy (conditional on the channel realizations $\{E_{k,l},\mathbf{H}_{k,l}\}_{l=1}^{L}$) of each estimate to unity and obtains the normalized estimates $\hat{s}^{\mathrm{U}}_{k,l,m}$. Finally, $\mathcal{R}_k$ passes the normalized estimates through output linear-beamforming (column) vectors $\mathbf{v}_{k,l,m}\in\mathbb{C}^N$ (which are designed to exploit the knowledge of the forward CSI $\{F_{k,l},\mathbf{G}_{k,l}\}_{l=1}^{L}$) to produce its transmit signal vector

$$\mathbf{t}_k=\frac{\sqrt{P_{\mathcal{R}}}}{LM}\sum_{p=1}^{L}\sum_{q=1}^{M}\frac{\mathbf{v}_{k,p,q}}{\|\mathbf{v}_{k,p,q}\|}\,\hat{s}^{\mathrm{U}}_{k,p,q},$$

concurrently ensuring that the transmit power constraint is satisfied. Hence under LDMRB, it follows that the $m$-th element of the signal vector $\mathbf{y}_l$ received at $\mathcal{D}_l$ is given by

$$y_{l,m}=\sum_{k=1}^{K}\frac{\sqrt{F_{k,l}P_{\mathcal{R}}}}{LM}\sum_{p=1}^{L}\sum_{q=1}^{M}\frac{\mathbf{g}_{k,l,m}\,\mathbf{v}_{k,p,q}}{\|\mathbf{v}_{k,p,q}\|}\,\hat{s}^{\mathrm{U}}_{k,p,q}+z_{l,m},$$

where $\mathbf{g}_{k,p,q}$ is the $q$-th row of $\mathbf{G}_{k,p}$. We list the input and output linear relay beamforming matrices $\{\mathbf{U}_k\}_{k=1}^{K}$ and $\{\mathbf{V}_k\}_{k=1}^{K}$ based on the MF, ZF and L-MMSE algorithms in Table 1. Here, the row vector $\mathbf{u}_{k,l,m}\in\mathbb{C}^N$ is the $((l-1)M+m)$-th row of $\mathbf{U}_k\in\mathbb{C}^{LM\times N}$ and the column vector $\mathbf{v}_{k,l,m}\in\mathbb{C}^N$ is the $((l-1)M+m)$-th column of $\mathbf{V}_k\in\mathbb{C}^{N\times LM}$.

**Spectral Efficiency vs. $E_b/N_0$.** The following theorem provides our main result on the power-bandwidth tradeoff in dense MRNs with practical LDMRB schemes.

**Theorem 2:** *The asymptotic power-bandwidth tradeoff for dense MRNs under LDMRB schemes, as the number of relay terminals tends to infinity, can be characterized as follows:*

**Low $E_b/N_0$ regime.** *In the limit of large $K$, MRN power-bandwidth tradeoff for LDMRB schemes under MF, ZF and L-MMSE algorithms almost surely converges to the deterministic relationship*

$$\frac{E_b}{N_0}(\mathsf{C})=\sqrt{\frac{L^3M^3}{\Theta_1^2K}\frac{2^{2\mathsf{C}(LM)^{-1}}-1}{\mathsf{C}^2}}+o\left(\frac{1}{\sqrt{K}}\right), \tag{7}$$

*where* $\Theta_1=\mathbb{E}\left[\sqrt{E_{k,l}F_{k,l}X_{k,l,m}Y_{k,l,m}}\right]$ *and fading-dependent random variables* $X_{k,l,m}$ *and* $Y_{k,l,m}$ *(independent across* $k$*) follow the* $\Gamma(N)$ *probability distribution* $p(\gamma)=(\gamma^{N-1}e^{-\gamma})/(N-1)!$ *for the MF and L-MMSE algorithms and* $\Gamma(N-LM+1)$ *distribution for the ZF algorithm. All LDMRB schemes achieve the minimum energy per bit at a finite spectral efficiency given by* $\mathsf{C}^*\approx 1.15\,LM$ *and consequently*

$$\frac{E_b}{N_0}_{\min}^{\mathrm{LDMRB}}\approx\sqrt{\frac{2.97LM}{\Theta_1^2K}}+o\left(\frac{1}{\sqrt{K}}\right),\ K\rightarrow\infty. \tag{8}$$

**High $E_b/N_0$ regime.** *In the limit of large $K$, MRN power-bandwidth tradeoff for LDMRB schemes under ZF and L-MMSE algorithms almost surely converges to the deterministic relationship*

$$\frac{E_b}{N_0}(\mathsf{C})=\frac{2^{2\mathsf{C}(LM)^{-1}}}{2\mathsf{C}}\,\frac{LM}{K\,\Theta_3^2}\left(\sqrt{\Theta_2}+\sqrt{L\,M}\right)^2+o\left(\frac{1}{K}\right), \tag{9}$$

[9] $\xrightarrow{\text{w.p.1}}$ denotes convergence with probability 1 (also known as almost sure convergence) [33].

| Channel Description | MF | ZF | L-MMSE |
| --- | --- | --- | --- |
| $\{\mathcal{S}_l\}_{l=1}^{L} \to \mathcal{R}_k$ Links $: \mathbf{H}_k = \begin{bmatrix} \sqrt{E_{k,1}}\,\mathbf{H}_{k,1}^T \\ \vdots \\ \sqrt{E_{k,L}}\,\mathbf{H}_{k,L}^T \end{bmatrix}^T$ | $\mathbf{U}_k = \mathbf{H}_k^H$ | $\mathbf{U}_k = (\mathbf{H}_k^H\mathbf{H}_k)^{-1}\mathbf{H}_k^H$ | $\mathbf{U}_k = (\frac{MN_0B}{P_\mathcal{S}}\mathbf{I} + \mathbf{H}_k^H\mathbf{H}_k)^{-1}\mathbf{H}_k^H$ |
| $\mathcal{R}_k \to \{\mathcal{D}_l\}_{l=1}^{L}$ Links $: \mathbf{G}_k = \begin{bmatrix} \sqrt{F_{k,1}}\,\mathbf{G}_{k,1} \\ \vdots \\ \sqrt{F_{k,L}}\,\mathbf{G}_{k,L} \end{bmatrix}$ | $\mathbf{V}_k = \mathbf{G}_k^H$ | $\mathbf{V}_k = \mathbf{G}_k^H(\mathbf{G}_k\mathbf{G}_k^H)^{-1}$ | $\mathbf{V}_k = \mathbf{G}_k^H(\frac{NN_0B}{P_\mathcal{R}}\mathbf{I} + \mathbf{G}_k\mathbf{G}_k^H)^{-1}$ |

Table 1. Practical LDMRB Schemes for multi-user MRNs.

*where* $\Theta_2 = \mathbb{E}\left[(F_{k,l}X_{k,l,m})/(E_{k,l}Y_{k,l,m})\right]$, $\Theta_3 = \mathbb{E}\left[\sqrt{F_{k,l}X_{k,l,m}}\right]$ *and fading-dependent random variables* $X_{k,l,m}$ *and* $Y_{k,l,m}$ *(independent across* $k$*) follow the* $\Gamma(N - LM + 1)$ *probability distribution. This power-bandwidth tradeoff leads to*

$$\frac{E_b}{N_0}_{\mathrm{imp}}^{\mathrm{ZF,L-MMSE}} = \frac{LM}{2K\Theta_3^2}\left(\sqrt{\Theta_2} + \sqrt{LM}\right)^2 + o\left(\frac{1}{K}\right)$$
$$S_\infty^{\mathrm{ZF,L-MMSE}} = \frac{LM}{2}, \qquad K \to \infty. \tag{10}$$

*The MRN operates in the interference-limited regime under MF-LDMRB and* $\mathsf{C}^{\mathrm{MF}}$ *converges to a fixed constant (which scales like* $\log(K)$*) as* $E_b/N_0 \to \infty$*; leading to* $S_\infty^{\mathrm{MF}} = 0$.

**Proof:** In the presence of full spatial multiplexing and horizontal encoding/decoding as discussed in Section II, each spatial stream at the destination terminals is decoded with no attempt to exploit the knowledge of the codebooks of the $LM - 1$ interfering streams (i.e., independent decoding); and instead this interference is treated as Gaussian noise. Consequently, the spectral efficiency of multi-user MRN can be expressed as

$$\mathsf{C}^{\mathrm{MRN}} = \frac{1}{2}\sum_{l=1}^{L}\sum_{m=1}^{M}\mathbb{E}_{\{\mathbf{H}_{k,l},\mathbf{G}_{k,l}\}}\left[\log_2\left(1 + \mathsf{SIR}_{l,m}\right)\right], \tag{11}$$

where $\mathsf{SIR}_{l,m}$ is the received signal-to-interference-plus-noise ratio (SIR) corresponding to spatial stream $s_{l,m}$ at terminal $\mathcal{D}_l$. The rest of the proof involves the analysis of low and high $E_b/N_0$ asymptotic behavior of (11) as a function of $\mathsf{SIR}_{l,m}$ in the limit of large $K$ for LDMRB schemes under the MF, ZF and L-MMSE algorithms. Here we present the detailed power-bandwidth tradeoff analysis for the ZF-based and MF-based LDMRB schemes in the high and low $E_b/N_0$ regimes. The LDMRB performance under the L-MMSE algorithm is identical to that of the ZF algorithm in the high $E_b/N_0$ regime and to that of the MF algorithm in the low $E_b/N_0$ regime.

*A. Proof for the ZF-LDMRB Scheme:*

It is easy to show that (see [35]) for the ZF-LDMRB scheme, the signal received at the $m$-th antenna of destination terminal $\mathcal{D}_l$ corresponding to spatial stream $s_{l,m}$ is given by

$$y_{l,m}^{\mathrm{ZF}} = \left(\sum_{k=1}^{K} d_{k,l,m}\right)s_{l,m} + \sum_{k=1}^{K} d_{k,l,m}\,\widetilde{n}_{k,l,m} + z_{l,m}, \tag{12}$$

where

$$d_{k,l,m} = \sqrt{\frac{P_\mathcal{R}F_{k,l}X_{k,l,m}}{L^2M^2\left(\frac{P_\mathcal{S}}{M} + \left(\frac{E_{k,l}Y_{k,l,m}}{N_0B}\right)^{-1}\right)}}, \tag{13}$$

and $\widetilde{n}_{k,l,m}$ denotes the $m$-th element of the vector $\widetilde{\mathbf{n}}_{k,l} = (E_{k,l})^{-1/2}\mathbf{D}_{k,l}\mathbf{n}_k$ and fading-dependent random variables $X_{k,l,m}$ and $Y_{k,l,m}$ follow the $\Gamma(N - LM + 1)$ probability distribution. The matrices $\{\mathbf{D}_{k,l}\}$ are obtained by letting $\mathbf{F}_k = \begin{bmatrix} \mathbf{H}_{k,1} & \cdots & \mathbf{H}_{k,L} \end{bmatrix}$, and setting $\mathbf{F}_k^\dagger = (\mathbf{F}_k^H\mathbf{F}_k)^{-1}\mathbf{F}_k^H$, which leads to

$$\mathbf{F}_k^\dagger = \begin{bmatrix} \mathbf{D}_{k,1} \\ \vdots \\ \mathbf{D}_{k,L} \end{bmatrix},$$

where each $\mathbf{D}_{k,l} \in \mathbb{C}^{M\times N}$. As a result, the ZF-LDMRB scheme decouples the effective channels between source-destination pairs $\{\mathcal{S}_l \to \mathcal{D}_l\}_{l=1}^{L}$ into $LM$ parallel spatial channels. From (12)-(13), we compute $\mathsf{SIR}_{l,m}$ as given in (14). We shall now continue our analysis by investigating the low and high $E_b/N_0$ regimes separately:

*Low $E_b/N_0$ regime:* If $\mathsf{SNR} \ll 1$, then $\mathsf{SIR}_{l,m}^{\mathrm{ZF}}$ in (14) simplifies to (15). Under the assumption that $\{E_{k,l}\}$ and $\{F_{k,l}\}$ are positive and bounded, we obtain

$$\sum_{k=1}^{K}\frac{\sqrt{E_{k,l}F_{k,l}X_{k,l,m}Y_{k,l,m}}}{K} - \sum_{k=1}^{K}\frac{\Theta_1}{K} \xrightarrow{\text{w.p.1}} 0$$

as $K \to \infty$, yielding (based on Theorems 1.8.D and 1.7 in [32])

$$\mathsf{SIR}_{l,m}^{\mathrm{ZF}} \xrightarrow{\text{w.p.1}} \frac{P_\mathcal{S}}{N_0\,B}\,\frac{P_\mathcal{R}}{N_0\,B}\,\frac{K^2}{L^2\,M^3}\,\Theta_1^2 + o(K). \tag{16}$$

Letting $\beta = P_\mathcal{R}/P_\mathcal{S}$, we find that SIR-maximizing power allocation (for fixed SNR) is achieved with $\beta^* = L/K$ resulting in (for $\mathsf{SNR} \ll 1$)

$$\mathsf{SIR}_{l,m}^{\mathrm{ZF}} \xrightarrow{\text{w.p.1}} \mathsf{SNR}^2\left(\frac{K\Theta_1^2}{L^3M^3} + o(K)\right), \tag{17}$$
$$\mathsf{C}^{\mathrm{ZF}} \xrightarrow{\text{w.p.1}} \frac{LM}{2}\log_2\left(1 + \mathsf{SNR}^2\left(\frac{K\Theta_1^2}{L^3M^3} + o(K)\right)\right) \tag{18}$$

$$\mathsf{SIR}_{l,m}^{\mathrm{ZF}} = \frac{P_{\mathcal{S}} K^2 \left( \frac{1}{K} \sum_{k=1}^{K} \sqrt{P_{\mathcal{R}} F_{k,l} X_{k,l,m} \left( L^2 M^2 \left( \frac{P_{\mathcal{S}}}{M} + \left( \frac{E_{k,l} Y_{k,l,m}}{N_0 B} \right)^{-1} \right) \right)^{-1}} \right)^2}{M N_0 B \left( 1 + K \frac{1}{K} \sum_{k=1}^{K} P_{\mathcal{R}} F_{k,l} X_{k,l,m} \left( L^2 M^2 \left( \frac{E_{k,l} P_{\mathcal{S}}}{M} Y_{k,l,m} + N_0 B \right) \right)^{-1} \right)} \quad (14)$$

$$\mathsf{SIR}_{l,m}^{\mathrm{ZF}} = \frac{P_{\mathcal{S}}}{N_0\, B} \frac{P_{\mathcal{R}}}{N_0\, B} \frac{K^2}{L^2\, M^3} \left( \frac{1}{K} \sum_{k=1}^{K} \sqrt{E_{k,l} F_{k,l} X_{k,l,m} Y_{k,l,m}} \right)^2 \quad (15)$$

Substituting $\mathsf{SNR} = \mathsf{C}\frac{E_b}{N_0}$ into (18) and solving for $\frac{E_b}{N_0}$, we obtain the result in (7). The rest of the proof follows from the strict convexity of $(2^{2\mathsf{C}(LM)^{-1}} - 1)/\mathsf{C}^2$ in $\mathsf{C}$ for all $\mathsf{C} \geq 0$.

*High $E_b/N_0$ regime:* If $\mathsf{SNR} \gg 1$, then $\mathsf{SIR}_{l,m}^{\mathrm{ZF}}$ in (14) simplifies to

$$\mathsf{SIR}_{l,m}^{\mathrm{ZF}} = \frac{P_{\mathcal{S}} K^2 \left( \frac{1}{K} \sum_{k=1}^{K} \sqrt{\frac{P_{\mathcal{R}} F_{k,l} X_{k,l,m}}{L^2 M P_{\mathcal{S}}}} \right)^2}{M\, N_0\, B \left( 1 + K \frac{1}{K} \sum_{k=1}^{K} \frac{P_{\mathcal{R}} F_{k,l}\, X_{k,l,m}}{L^2 M E_{k,l} Y_{k,l,m} P_{\mathcal{S}}} \right)}.$$

It follows from Theorem 1.8.D in [32] that as $K \to \infty$

$$\frac{1}{K} \sum_{k=1}^{K} \frac{F_{k,l} X_{k,l,m}}{E_{k,l} Y_{k,l,m}} - \sum_{k=1}^{K} \frac{\Theta_2}{K} \xrightarrow{\text{w.p.1}} 0.$$

$$\sum_{k=1}^{K} \frac{\sqrt{F_{k,l} X_{k,l,m}}}{K} - \sum_{k=1}^{K} \frac{\Theta_3}{K} \xrightarrow{\text{w.p.1}} 0,$$

Now applying Theorem 1.7 in [32], we obtain

$$\mathsf{SIR}_{l,m}^{\mathrm{ZF}} \xrightarrow{\text{w.p.1}} \frac{K^2 \Theta_3^2}{N_0 B \left( \frac{L^2 M^2}{P_{\mathcal{R}}} + \frac{K M}{P_{\mathcal{S}}} \Theta_2 \right)} + o(K).$$

Letting $\beta = P_{\mathcal{R}}/P_{\mathcal{S}}$, the SIR-maximizing power allocation (for fixed SNR) is achieved with $\beta^* = \sqrt{L^3 M/(K^2 \Theta_2)}$ resulting in ($\mathsf{SNR} \gg 1$)

$$\mathsf{SIR}_{l,m}^{\mathrm{ZF}} \xrightarrow{\text{w.p.1}} \frac{2K\mathsf{SNR}}{LM} \frac{\Theta_3^2}{\left( \sqrt{\Theta_2} + \sqrt{LM} \right)^2} + o(K). \quad (19)$$

Now substituting (19) into (11), we obtain

$$\mathsf{C}^{\mathrm{ZF}} \xrightarrow{\text{w.p.1}} \frac{LM}{2} \log_2 \left( \frac{2K\mathsf{SNR}}{LM} \frac{\Theta_3^2}{\left( \sqrt{\Theta_2} + \sqrt{LM} \right)^2} + o(K) \right). \quad (20)$$

Applying (2)-(3) to $\mathsf{C}^{\mathrm{ZF}}$ in (20), we obtain the high $E_b/N_0$ power-bandwidth tradeoff relationships in (10).

*B. Proof for the MF-LDMRB Scheme:*

When the MF-LDMRB scheme is employed, terminal $\mathcal{R}_k$ correlates its received signal vector $\mathbf{r}_k$ with each of the spatial signature vectors $\mathbf{h}_{k,l,m}$ ($m$-th column of $\mathbf{H}_{k,l}$) to yield

$$\hat{s}_{k,l,m}^{\mathrm{MF}} = \sqrt{E_{k,l}}\, \|\mathbf{h}_{k,l,m}\|^2\, s_{l,m} + \sum_{(p,q)\neq(l,m)} \sqrt{E_{k,p}}\, \mathbf{h}_{k,l,m}^H \mathbf{h}_{k,p,q}\, s_{p,q} + \mathbf{h}_{k,l,m}^H\, \mathbf{n}_k,$$

as the MF estimate for $s_{l,m}$. After normalizing the average energy of the MF estimates (conditional on the channel realizations $\{E_{k,l}, \mathbf{H}_{k,l}\}_{l=1}^{L}$) to unity, the matched filter output is given by (21). Next, the relay terminal $\mathcal{R}_k$ pre-matches its forward channels to ensure that the intended signal components add coherently at their corresponding destination terminals, while satisfying its transmit power constraint, to produce the transmit signal vector

$$\mathbf{t}_k = \frac{\sqrt{P_{\mathcal{R}}}}{LM} \sum_{p=1}^{L} \sum_{q=1}^{M} \frac{\mathbf{g}_{k,p,q}^H}{\|\mathbf{g}_{k,p,q}\|}\, \hat{s}_{k,p,q}^{\mathrm{U,MF}},$$

where $\mathbf{g}_{k,p,q}$ is the $q$-th row of $\mathbf{G}_{k,p}$ and it follows that

$$y_{l,m}^{\mathrm{MF}} = \sum_{k=1}^{K} \frac{\sqrt{F_{k,l} P_{\mathcal{R}}}}{LM} \sum_{p=1}^{L} \sum_{q=1}^{M} \frac{\mathbf{g}_{k,l,m}\, \mathbf{g}_{k,p,q}^H}{\|\mathbf{g}_{k,p,q}\|}\, \hat{s}_{k,p,q}^{\mathrm{U,MF}} + z_{l,m}. \quad (22)$$

We shall now continue our analysis by investigating low and high $E_b/N_0$ regimes separately:

*Low $E_b/N_0$ regime:* Assuming that the system operates in the power-limited low SNR ($\mathsf{SNR} \ll 1$) regime, the noise power dominates over the signal and interference powers for the received signals at the relay and destination terminals. Consequently, the loss in the signal-to-interference-plus-noise ratio at each destination antenna due to the MF-based relays' incapability of interference cancellation is negligible. Hence, in the low $E_b/N_0$ regime, the expression for the received signal at the destination under MF relaying in (22) can be simplified as

$$y_{l,m}^{\mathrm{MF}} = \sum_{k=1}^{K} \frac{1}{LM} \sqrt{\frac{P_{\mathcal{R}} E_{k,l} F_{k,l}}{N_0 B}} ||\mathbf{h}_{k,l,m}|| ||\mathbf{g}_{k,l,m}|| s_{l,m} + z_{l,m}.$$

In this setting, $\mathsf{SIR}_{l,m}^{\mathrm{MF}}$ over each stream is given by (23), where $X_{k,l,m}$ and $Y_{k,l,m}$ follow the $\Gamma(N)$ distribution (note that the distribution of $X_{k,l,m}$ and $Y_{k,l,m}$ is different from the ZF case). Observing the similarity of the expression in (23) to (15), the rest of the proof is identical to the low $E_b/N_0$ analysis of the ZF-LDMRB scheme. We apply the same steps as in the proof of (17) to obtain (for $\mathsf{SNR} \ll 1$)

$$\mathsf{SIR}_{l,m}^{\mathrm{MF}} \xrightarrow{\text{w.p.1}} \mathsf{SNR}^2 \left( \frac{K \Theta_1^2}{L^3 M^3} + o(K) \right), \quad (24)$$

$$\mathsf{C}^{\mathrm{MF}} \xrightarrow{\text{w.p.1}} \frac{LM}{2} \log_2 \left( 1 + \mathsf{SNR}^2 \left( \frac{K \Theta_1^2}{L^3 M^3} + o(K) \right) \right), \quad (25)$$

$$\hat{s}_{k,l,m}^{\mathrm{U,MF}} = \frac{\sqrt{E_{k,l}}\, \|\mathbf{h}_{k,l,m}\|^2 s_{l,m} + \sum_{(p,q)\neq(l,m)} \sqrt{E_{k,p}}\, \mathbf{h}_{k,l,m}^H \mathbf{h}_{k,p,q}\, s_{p,q} + \mathbf{h}_{k,l,m}^H\, \mathbf{n}_k}{\sqrt{E_{k,l}\, \|\mathbf{h}_{k,l,m}\|^4 \frac{P_{\mathcal{S}}}{M} + \sum_{(p,q)\neq(l,m)} E_{k,p} |\mathbf{h}_{k,l,m}^H \mathbf{h}_{k,p,q}|^2 \frac{P_{\mathcal{S}}}{M} + \|\mathbf{h}_{k,l,m}\|^2 N_0 B}} \tag{21}$$

$$\mathsf{SIR}_{l,m}^{\mathrm{MF}} = \frac{P_{\mathcal{S}}}{N_0\, B} \frac{P_{\mathcal{R}}}{N_0\, B} \frac{K^2}{L^2\, M^3} \left( \frac{1}{K} \sum_{k=1}^{K} \sqrt{E_{k,l} F_{k,l} X_{k,l,m} Y_{k,l,m}} \right)^2 \tag{23}$$

which finally leads to the result in (7).

*High $E_b/N_0$ regime:* Due to the tedious nature of the analysis of the MF-LDMRB scheme in the high SNR regime, here we shall only provide a non-rigorous argument to justify why this scheme leads to interference-limited network behavior. Assuming that the system operates in the high SNR regime ($\mathsf{SNR} \gg 1$), the signal and interference powers dominate over the noise power for the received signals at the relays and destination. Due to the fact that $P_{\mathcal{S}} \gg N_0 B$, the majority of the transmitted signal at the relay terminals is composed of signal and interference components and therefore the amplification of noise at the relays due to linear processing contributes negligibly to the SIR at the destination for all multiplexed streams. Thus at high SNR, the spectral efficiency of the MRN under MF-LDMRB is of the form

$$\mathsf{C}^{\mathrm{MF}} = \frac{LM}{2} \mathbb{E}\left[ \log_2 \left( 1 + \frac{P_{\mathcal{R}} f_{l,m}^{\mathrm{sig}}}{P_{\mathcal{R}} f_{l,m}^{\mathrm{int}} + N_0 B f_{l,m}^{\mathrm{noise}}} \right) \right],$$

where the SIR of each stream, $\mathsf{SIR}_{l,m}^{\mathrm{MF}}$, is determined by the positive-valued functions $f_{l,m}^{\mathrm{sig}}$, $f_{l,m}^{\mathrm{int}}$ and $f_{l,m}^{\mathrm{noise}}$, which specify the dependence of the powers of the signal, interference and noise components, respectively, (for the stream $s_{l,m}$) on the set of MRN channel realizations $\{E_{k,l}, F_{k,l}, \mathbf{H}_{k,l}, \mathbf{G}_{k,l}\}$. Since $P_{\mathcal{R}} \gg N_0 B$, the signal and interference powers dominate the power due to additive noise at each destination. Furthermore, since the signal and interference components grow at the same rate with respect to SNR, as $\mathsf{SNR} \to \infty$, the SIR of each stream will no longer be proportional to SNR (which is not true for ZF and L-MMSE LDMRB due to their ability to suppress interference) resulting in *interference-limitedness* and the convergence of $\mathsf{C}^{\mathrm{MF}}$ to a fixed limit independent of SNR. Fixing $K$ to be large but finite and letting $\mathsf{SNR} \to \infty$, we have

$$\begin{aligned} \lim_{\mathsf{SNR}\to\infty} \frac{E_b}{N_0}^{\mathrm{MF}} &= \lim_{\mathsf{SNR}\to\infty} \frac{\mathsf{SNR}}{C^{\mathrm{MF}}(\mathsf{SNR})} \\ &= \lim_{\mathsf{SNR}\to\infty} \frac{\mathsf{SNR}}{\mathrm{constant}} \to \infty, \end{aligned}$$

and consequently $S_\infty^{\mathrm{MF}} = 0$. On the other hand, for fixed SNR, from the capacity scaling analysis of MF-LDMRB in [6], we know that

$$\mathsf{C}^{\mathrm{MF}} = \frac{LM}{2} \log_2(K) + o\left(\log_2(K)\right), \quad K \to \infty,$$

since the signal power grows faster than the interference power as $K \to \infty$. Thus while the optimal spectral efficiency scaling is maintained by MF-LDMRB, the energy efficiency performance becomes poor due to relays' inability to suppress interference. □

**Interpretation of Theorem 2.** The key results in (17), (19) and (24) give a complete picture in terms of how LDMRB impacts the SIR statistics at the destination terminal in the low and high $E_b/N_0$ regimes. We emphasize that the conclusions related to MF-LDMRB in the low $E_b/N_0$ regime and those related to ZF-LDMRB in the high $E_b/N_0$ regime apply for the L-MMSE algorithm (L-MMSE converges to ZF as $\mathsf{SNR} \to \infty$ and to MF as $\mathsf{SNR} \to 0$), and therefore our analysis has provided insights for the energy efficiency and spectral efficiency of all three (MF, ZF and L-MMSE) different LDMRB schemes. We make the following observations:

*Remark 1:* We observe from (17), (19) and (24) that $\mathsf{SIR}_{l,m}$ scales *linearly* in the number of relay terminals, $K$, providing higher energy efficiency.[10] We emphasize that the linear scaling of $\mathsf{SIR}_{l,m}$ in the number of relay terminals $K$, is maintained independent of SNR (i.e., valid for both low and high SNR). This can be interpreted as *distributed energy efficiency gain*, since it is realized without requiring any cooperation among the relay terminals.

*Remark 2:* The SIR scaling results in (17), (19) and (24) have been key toward proving the scaling results on $E_b/N_0$ given by (8)-(10). Our asymptotic analysis shows that $E_b/N_0$ reduces like $K^{-1/2}$ in the low $E_b/N_0$ regime for LDMRB under the MF, ZF and L-MMSE algorithms and like $K^{-1}$ in the high $E_b/N_0$ regime for the ZF and L-MMSE algorithms. Thus, ZF and L-MMSE algorithms achieve *optimal energy scaling* (in $K$) for high $E_b/N_0$ (the fact that $K^{-1}$ is the best-possible energy scaling was established in Theorem 1 based on the cut-set upper bound). Furthermore, unlike MF, the spectral efficiency of the ZF and L-MMSE algorithms grows without bound with $E_b/N_0$ due to their interference cancellation capability and achieves the *optimal high-SNR slope* (as in the cutset bound) of $S_\infty = LM/2$. In the high $E_b/N_0$ regime, Theorem 2 shows that for fixed $K$, the growth of SNR does not lead to an increase in spectral efficiency for MF-LDMRB; and the spectral efficiency saturates to a fixed value (from [6], we know that this fixed spectral efficiency value scales like $\log_2(K)$), leading to $S_\infty = 0$ and a poor power-bandwidth tradeoff due to the interference-limited network behavior.

*Remark 3:* We observe from the almost sure convergence results in (17), (19) and (24) on the SIR statistics that LDMRB schemes realize *cooperative diversity gain* [36]-[37] arising from the deterministic scaling behavior of $\mathsf{SIR}_{l,m}$ in $K$. Hence in the limit of infinite number of relays, a Shannon capacity exists even for an MRN under the slow fading (non-ergodic) channel model [38] and thus our asymptotic results are valid without the ergodicity assumption on the channel statistics. This phenomenon of "relay ergodization" can be interpreted as a form of statistical averaging (over the spatial dimension created due to the assistance of multiple relay terminals) that ensures the convergence of the SIR statistics to a deterministic scaling behavior even if the fading processes affecting the individual relays are not ergodic. Even more importantly, the deterministic scaling behavior also suggests that the lack of CSI knowledge at the destination terminals does not degrade performance in the limit of infinite number of relay terminals.

[10]The fact that $\mathsf{SIR}_{l,m}$ scales linearly in $K$ for MF-LDMRB in the high $E_b/N_0$ regime has not been treated rigorously in this paper, a detailed analysis of this case can be found in [6].

*Remark 4:* Finally, we observe that all LDMRB schemes achieve the highest energy efficiency at a finite spectral efficiency. In other words, the most efficient power utilization under LDMRB is achieved at a *finite bandwidth* and there is *no power-bandwidth tradeoff* above a certain bandwidth. Additional bandwidth requires more power. A similar observation was made in [39] and [40] in the context of Gaussian parallel relay networks. The cause of this phenomenon is noise amplification, which significantly degrades performance at low SNRs when the MRN becomes noise-limited. We find that the ZF algorithm performs worse than the MF and L-MMSE algorithms in the low $E_b/N_0$ regime because of its inherent inability of noise suppression (the loss in SIR experienced by the ZF algorithm in the low $E_b/N_0$ regime can be explained from our analysis; we have seen that $X_{k,l,m}$ and $Y_{k,l,m}$ follow the $\Gamma(N)$ distribution at low $E_b/N_0$ for the MF and L-MMSE algorithms, while these fading-dependent random variables follow the $\Gamma(N-LM+1)$ distribution for the ZF algorithm).

**Bursty signaling in the low SNR regime.** One solution to the problem of noise amplification in the low SNR regime is *bursty* transmission [41]. For the duty cycle parameter $\alpha \in [0,1]$, this means that the sources and relays transmit only $\alpha$ fraction of time over which they consume total power $P/\alpha$ and remain silent otherwise; and hence satisfying the average power constraints. The result of bursty transmission is that the network is forced to operate in the high SNR regime at the expense of lower spectral efficiency. This is achieved, for instance under the ZF-LDMRB scheme, through the adjustment of signal burstiness by choosing the duty cycle parameter $\alpha$ small enough so that the condition

$$\alpha \ll \min_{k,l,m} \frac{E_{k,l} Y_{k,l,m} P_{\mathcal{S}}}{M N_0 B}, \tag{26}$$

is satisfied, which ensures that the linear beamforming operations at the relay terminals are performed under high SNR conditions and thus the detrimental impact of noise amplification on energy efficiency is minimized.[11] With such bursty signaling, even though $\mathsf{SNR} \ll 1$, the SIR for each stream in (14) simplifies to (note the additional $\alpha$ term in the denominator)

$$\mathsf{SIR}^{\mathrm{ZF,bursty}}_{l,m} = \frac{P_{\mathcal{S}} K^2 \left( \frac{1}{K} \sum_{k=1}^{K} \sqrt{\frac{P_{\mathcal{R}} F_{k,l} X_{k,l,m}}{L^2 M P_{\mathcal{S}}}} \right)^2}{\alpha M N_0 B \left( 1 + K \frac{1}{K} \sum_{k=1}^{K} \frac{P_{\mathcal{R}} F_{k,l} X_{k,l,m}}{L^2 M E_{k,l} Y_{k,l,m} P_{\mathcal{S}}} \right)}$$

as in the high $E_b/N_0$ regime and the network spectral efficiency is computed as

$$\mathsf{C}^{\mathrm{ZF,bursty}} = \frac{\alpha}{2} \sum_{l=1}^{L} \sum_{i=1}^{M} \mathbb{E}\left[ \log_2 \left( 1 + \mathsf{SIR}^{\mathrm{ZF,bursty}}_{l,m} \right) \right].$$

Hence, the results of Theorem 2 in (9) can immediately be applied, with slight modifications, resulting in the power-bandwidth tradeoff relation

$$\frac{E_b}{N_0}(\mathsf{C}) = \frac{2^{2\mathsf{C}(\alpha LM)^{-1}}}{2\mathsf{C}\,(\alpha LM)^{-1}} \frac{\left( \sqrt{\Theta_2} + \sqrt{LM} \right)^2}{K\,\Theta_3^2} + o\left( \frac{1}{K} \right). \tag{27}$$

The energy efficiency and spectral efficiency performance can be quantified by applying (2)-(3) to (27) yielding

$$\begin{aligned} \frac{E_b}{N_0}^{\mathrm{ZF,bursty}}_{\mathrm{imp}} &= \frac{LM}{2K\Theta_3^2} \left( \sqrt{\Theta_2} + \sqrt{LM} \right)^2 + o\left( \frac{1}{K} \right) \\ S_\infty^{\mathrm{ZF,bursty}} &= \frac{\alpha LM}{2}, \qquad K \to \infty. \end{aligned}$$

As a result, we have shown that with sufficient amount of burstiness, the *optimal energy scaling* of $K^{-1}$ can be achieved with the ZF (as well as L-MMSE) LDMRB schemes[12], while the high SNR spectral efficiency slope scales down by the duty cycle factor $\alpha$. Thus, *burstiness trades off spectral efficiency for higher energy efficiency*. We remark that our result establishes the *asymptotic optimality of LDMRB schemes* in the sense that with proper signaling they can alternately achieve the best possible (i.e., as in the cutset bound) energy efficiency scaling or the best possible spectral efficiency slope *for any SNR*. We also emphasize that our results proving that the energy scaling of $K^{-1}$ is achievable with LDMRB schemes enhances the result of previous work in [1], where the authors showed under an equivalent two-hop relay network model that linear relaying can only yield the energy scaling of $K^{-1/2}$.

## V. Numerical Results

The goal of this section is to support the conclusions of our theoretical analysis with numerical results. For the following examples, we set $E_{k,l}/(N_0 B) = F_{k,l}/(N_0 B) = 0\,\mathrm{dB}$.

**Example 1: SIR Statistics.** We consider an MRN with $L = 2$, $M = 1$ and $N = 2$ and analyze (based on Monte Carlo simulations) the SIR statistics for the LDMRB scheme based on the ZF algorithm and compare with the performance under direct transmissions. Direct transmission implies that the assistance from the relay terminals is not possible (i.e. $K = 0$), necessitating the two source terminals to transmit simultaneously over a common time and frequency resource to their intended destination terminals without the availability of relay interference cancellation mechanisms. For direct transmission, we assume that two source terminals share the total fixed average power $P$ equally (since there are no relay terminals involved, and there is a single time-slot for transmission) and thus we have $P_S = P/2$ and the network SNR is again given by $\mathsf{SNR} = P \,/\, (\, N_0 B \,)$. In this setting, the communication takes place over a fading interference channel [30] with single-user decoders at the destination terminals. Note that the direct transmission does not suffer from the $1/2$ capacity penalty that the LDMRB scheme incurs under the half-duplex two-hop transmission protocol. The channel distributions for the direct transmissions over the source-destination links are assumed to be identical to those over the MRN source-relay and relay-destination links (i.i.d. $\mathcal{CN}(0,1)$ statistics over all links). For fair comparison with LDMRB schemes, no transmit CSI is considered at the transmitters while the receivers possess perfect CSI. Denoting the overall channel gain (including path loss, shadowing and fading) between source $i \in \{1,2\}$ and destination $j \in \{1,2\}$ by $\xi_{i,j}$, the SIR for the stream corresponding to source-destination pair $j$ under direct transmission is

$$\mathsf{SIR}_j^{\mathrm{direct}} = \frac{|\xi_{j,j}|^2 \frac{P}{2}}{N_0 B + |\xi_{i,j}|^2 \frac{P}{2}}, \quad j \in \{1,2\},\, i \neq j.$$

[11] This implies that for block length $Q$, the number of symbol transmissions is given by $Q_{\mathrm{bursty}} = \lfloor \alpha Q \rfloor$ and that for strictly positive $\alpha$ that satisfies (26), as $Q \to \infty$, it is also true that $Q_{\mathrm{bursty}} \to \infty$, provided that $Q$ grows much faster than $K$ (since the growth of $K$ necessitates the choice of a smaller $\alpha$ under (26)). Thus the degrees of freedom (per codeword) necessary to cope with fading and additive noise are maintained and the Shannon capacity (ergodic mutual information) is achievable.

[12] The only necessary condition to achieve this optimal energy scaling is that $N \geq LM$ is satisfied so that the system does not become interference-limited at high SNR, which for instance would also apply to a single-user single-antenna relay network (where $L = M = N = 1$) under MF-LDMRB.

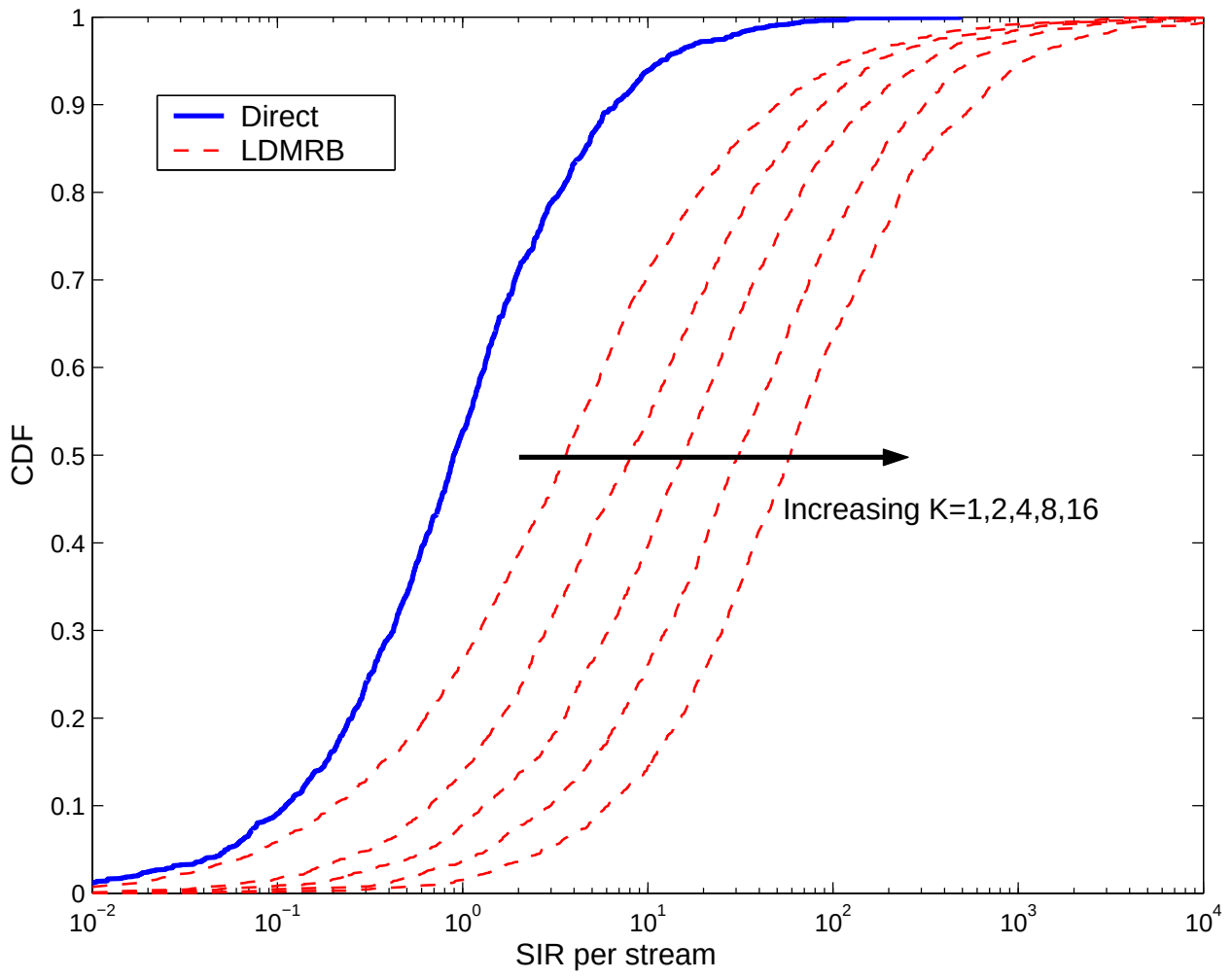

Fig. 3. CDF of SIR for direct transmission and (distributed) ZF-LDMRB for various values of $K$ at $\mathsf{SNR} = 20$ dB.

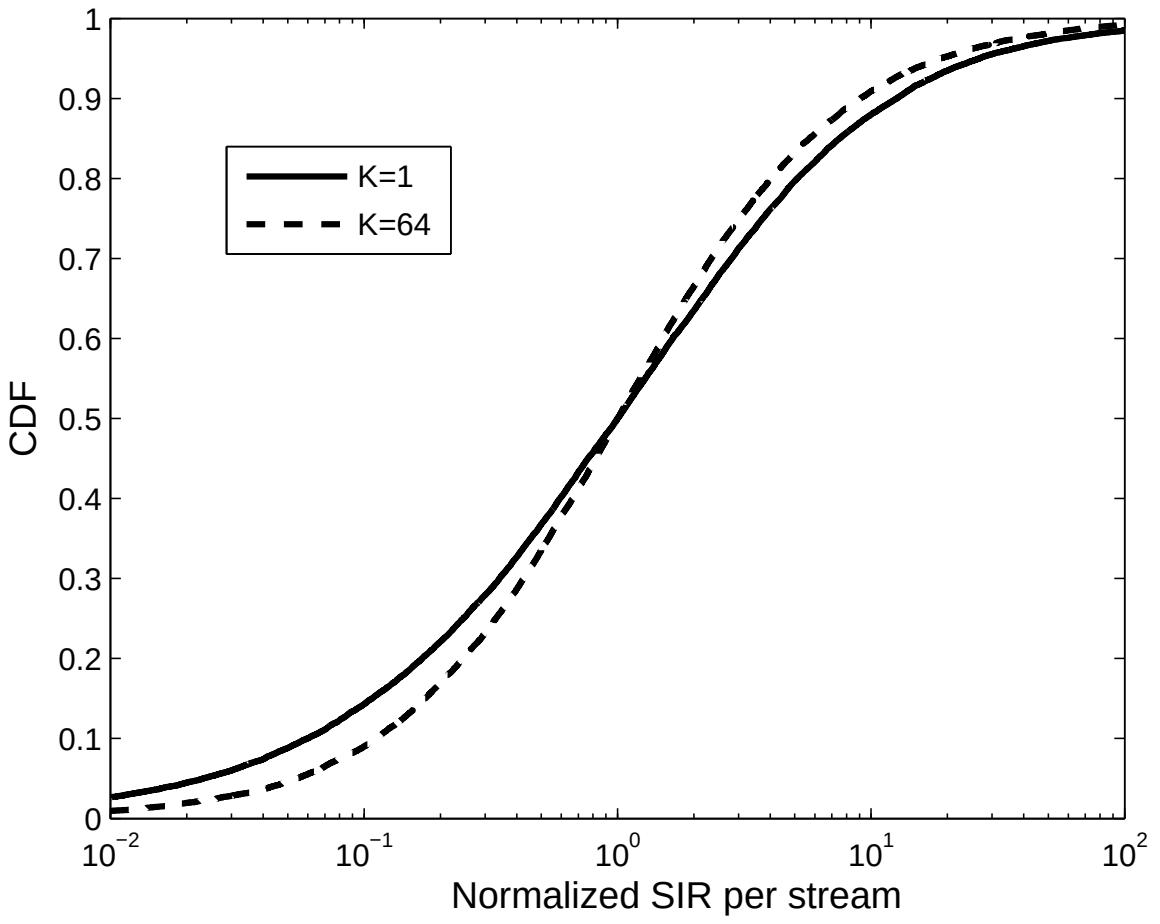

Fig. 4. CDF of normalized SIR (with respect to its median) per stream for ZF-LDMRB for $K = 1, 64$ at $\mathsf{SNR} = 20$ dB.

We set $\mathsf{SNR} = 20$ dB and plot the cumulative distribution function (CDF) of SIR for direct transmission and for the LDMRB scheme based on the ZF algorithm, and varying $K = 1, 2, 4, 8, 16$ in Fig. 3. As predicted by (19), we observe that the mean of SIR for LDMRB increases by $3$ dB for every doubling of $K$ due to the energy efficiency improvement proportional in the number of relay terminals. This verifies our analytical results in (17) and (19) indicating that the SIR of each multiplexed stream scales linearly in $K$ under ZF-LDMRB. We emphasize that these SIR scaling results have been key toward proving the $K^{-1/2}$ (at low SNR) and $K^{-1}$ (at high SNR) scaling results on $E_b/N_0$, and therefore this simulation result also serves toward verifying our energy scaling results given by (8)-(10) in Theorem 2. Furthermore, we note the huge improvement in SIR with respect to direct transmissions due to increased interference cancellation capability of the relay-assisted wireless network.

To illustrate the rate of convergence on the per-stream SIR statistics with respect to the growing number of relays, we plot in Fig. 4 the normalized SIR CDFs (normalization is performed by scaling the set of SIR realizations by its median) for the ZF-LDMRB scheme under the same assumptions for $K = 1, 64$. While the CDF of normalized per-stream SIR is tightening with increasing $K$, we observe that the convergence rate is slow and therefore we conclude that a large number of relay terminals (i.e., large $K$) is necessary to extract full merits of cooperative diversity gains.

**Example 2: MRN Power-Bandwidth Tradeoff.** We consider an MRN with $K = 10$, $L = 2$, $M = 1$ and $N = 2$ and numerically compute (based on Monte Carlo simulations) the average (i.e. ergodic) rates for the upper-limit based on the cutset bound, practical LDMRB schemes using MF, ZF and L-MMSE algorithms and direct transmission. We then use these average rates to compute spectral efficiency and energy efficiency quantified by $\mathsf{C} = R/B$ and $E_b/N_0 = \mathsf{SNR}/\mathsf{C}$, respectively, and repeat this process for various values of SNR to empirically obtain the power-bandwidth tradeoff curve for each scheme. We plot our numerical power-bandwidth tradeoff results in Fig. 5.

Our analytical results in (8)-(10) supported with the numerical results in Fig. 5 show that practical LDMRB schemes could yield significant power and bandwidth savings over direct transmissions. We observe that a significant portion of the set of energy efficiency and spectral efficiency pairs within the cutset outer bound (that is infeasible with direct transmission) is covered by practical LDMRB schemes. As our analytical results suggest, we see that the spectral efficiency of ZF and L-MMSE LDMRB grows without bound with $E_b/N_0$ due to the interference cancellation capability of these schemes and achieves the same high SNR slope as the cutset upper limit. Furthermore, this numerical exercise verifies our finding that in the high $E_b/N_0$ regime, the spectral efficiency of MF-LDMRB saturates to a fixed value leading to poor energy efficiency.

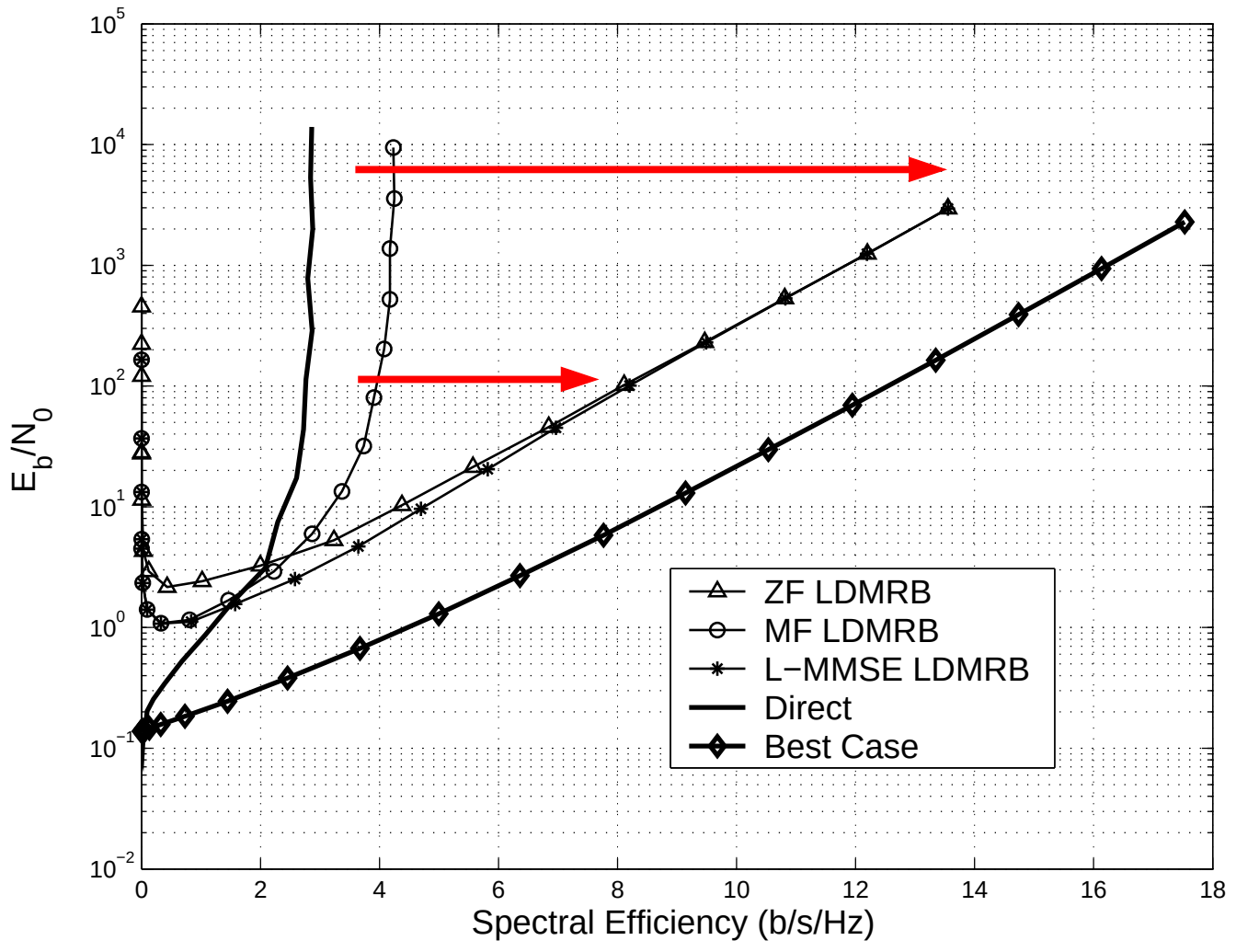

Fig. 5. MRN power-bandwidth tradeoff comparison: Upper-limit, practical LDMRB schemes and direct transmission.

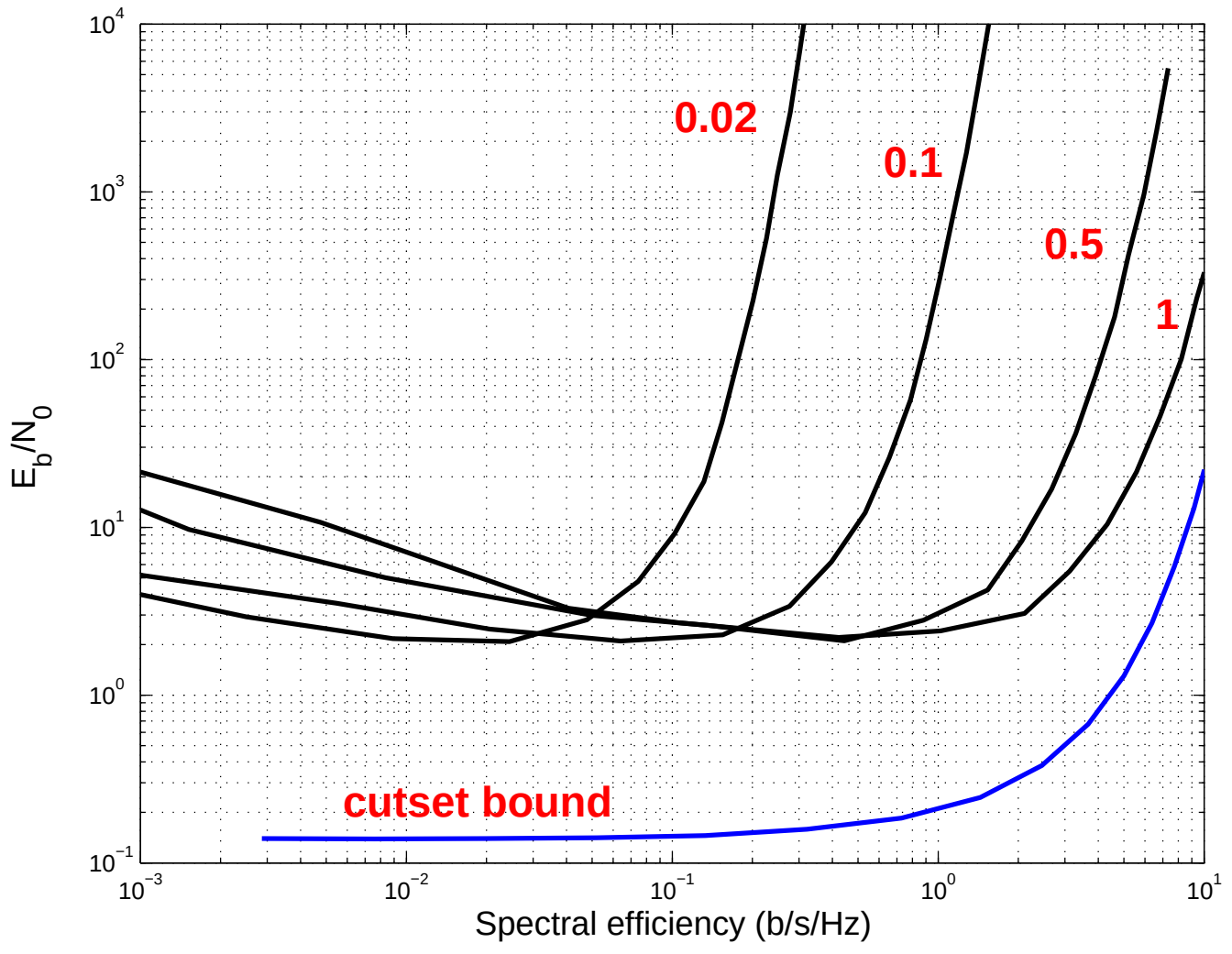

Fig. 6. Power-bandwidth tradeoff for the ZF-LDMRB scheme under bursty transmission for duty cycle parameters $\alpha = 0.02, 0.1, 0.5, 1$.

In Fig. 6, we plot power-bandwidth tradeoff under the ZF-LDMRB scheme (setting $K = 10$, $L = 2$, $M = 1$, $N = 2$) for duty cycle parameter values of $\alpha = 0.02, 0.1, 0.5, 1$. Clearly, we find that in the lower spectral efficiency (and hence lower SNR) regime, it is desirable to increase the level of burstiness by reducing the $\alpha$ parameter in order to achieve higher energy efficiency.

**Example 3: Enhancements from Multiple Antennas.** We consider an MRN under the L-MMSE LDMRB scheme with $K = 10$ and $L = 2$, and plot (based on Monte Carlo simulations) power-bandwidth tradeoff curves, obtained following the same methodology as Example 2, for different values of $M$ and $N$, to understand the impact of multi-antenna techniques at the source-destination pair and relay terminals on energy efficiency and spectral efficiency. From Fig. 7, it is clear that multiple antennas at the relay terminals improve energy efficiency (through the downward shift of the power-bandwidth tradeoff curve) while multiple antennas at the source-destination pairs improve spectral efficiency (through the improvement in the wideband slope and high SNR slope of the power-bandwidth tradeoff curve).

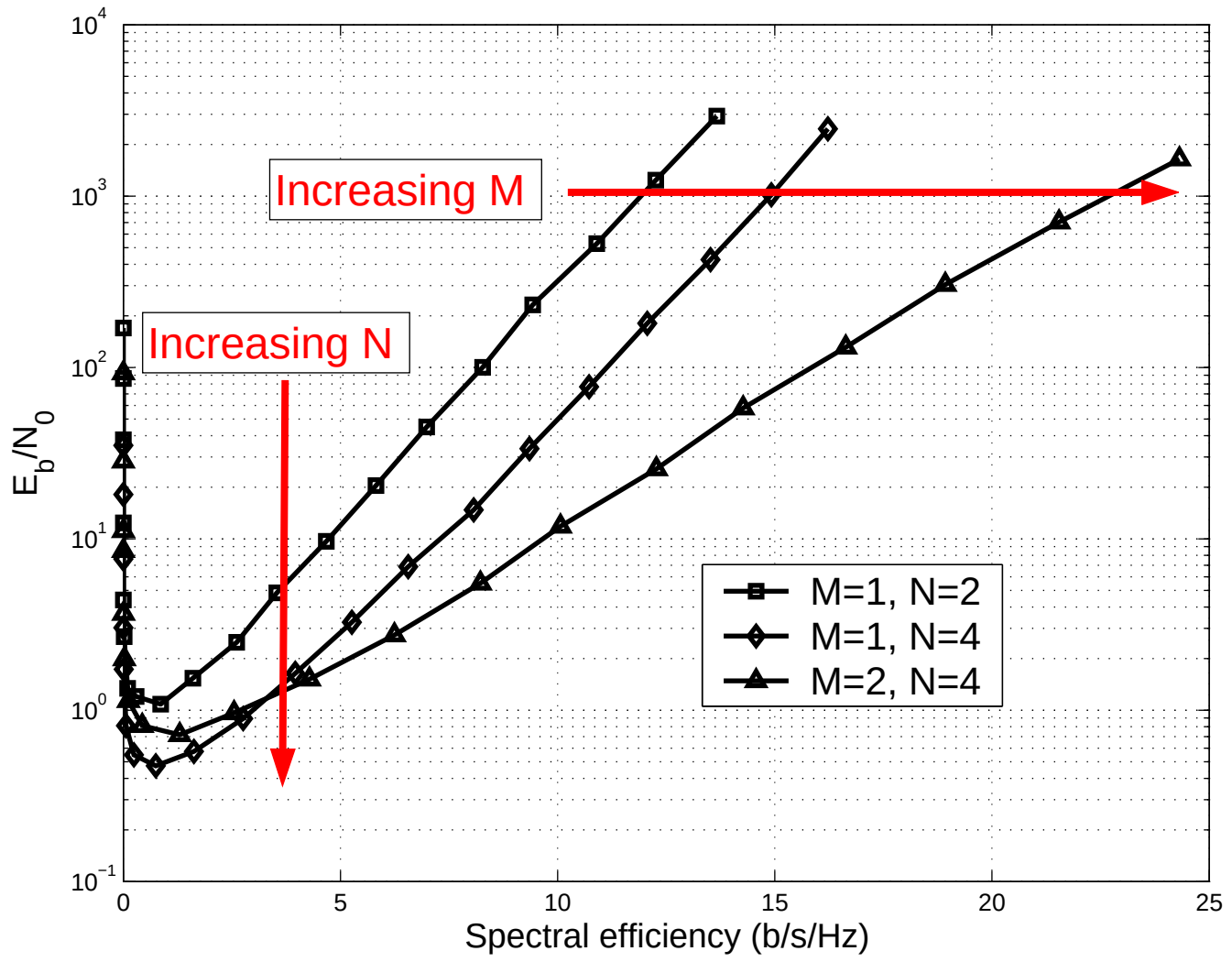

Fig. 7. MRN power-bandwidth tradeoff for the L-MMSE LDMRB scheme with varying number of antennas at the source-destination pair ($M$) and relay terminals ($N$).

## VI. Conclusions

As an additional leverage for supporting high data rates over next-generation wireless networks, we demonstrated how increasing density of wireless devices can be exploited by practical relay cooperation techniques to simultaneously improve energy efficiency and spectral efficiency. In particular, we designed low-complexity LDMRB schemes that exploit locally available channel state information (CSI) at each relay terminal to simultaneously convey multiple users' signals toward their intended destinations. Using Shannon-theoretic tools, we analyzed the power-bandwidth tradeoff for these techniques over a dense multi-user MRN model and demonstrated significant gains in terms of energy efficiency and spectral efficiency over direct transmissions. We established that in the limit of large number of relay terminals ($K \to \infty$), LDMRB schemes achieve asymptotically optimal power-bandwidth tradeoff at any SNR under bursty signaling capability; with the energy efficiency scaling like $K$. Finally, we verified our results through the numerical investigation of SIR statistics and power-bandwidth tradeoffs.

## Acknowledgment

The authors would like to thank Prof. Abbas El Gamal (Stanford) for beneficial discussions.

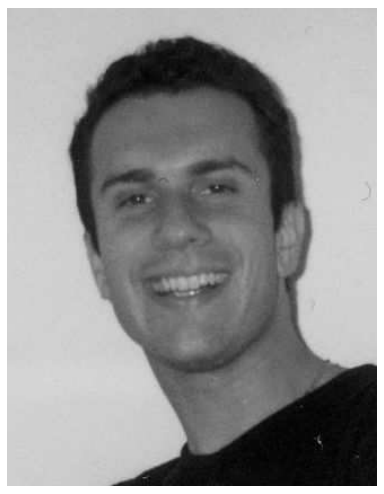

**Özgür Oyman** (S'02-M'05) was born in Istanbul, Turkey, on March 5, 1979. He received the B.S. (*summa cum laude*) degree in electrical engineering from Cornell University, Ithaca, NY, in 2000, and the M.S. and Ph.D. degrees in electrical engineering from Stanford University, Stanford, CA, in 2002 and 2005, respectively. Since September 2005, he has been working as a senior research scientist in the Corporate Technology Group at Intel Corporation, Santa Clara, CA.

During his doctoral studies, he was a member of the Smart Antennas Research Group within the Information Systems Laboratory at Stanford University. He was a visiting researcher at the Communication Theory Group at the Swiss Federal Institute of Technology (ETH) in Zürich, Switzerland in 2003. His prior industry experience includes work at Qualcomm (2001), Beceem Communications (2004) and Intel (2005). His research interests are in the applications of communication theory, information theory and mathematical statistics to wireless communications, with special emphasis on cross-layer design and system-level optimization of multiple-input multiple-output (MIMO) antenna systems and multihop/mesh/adhoc communication architectures.

Dr. Oyman was the recipient of the four-year Benchmark Stanford Graduate Fellowship. He is currently serving as the treasurer for the Santa Clara Valley Chapter of the IEEE Signal Processing Society. He is a member of Tau Beta Pi, Eta Kappa Nu and the IEEE.

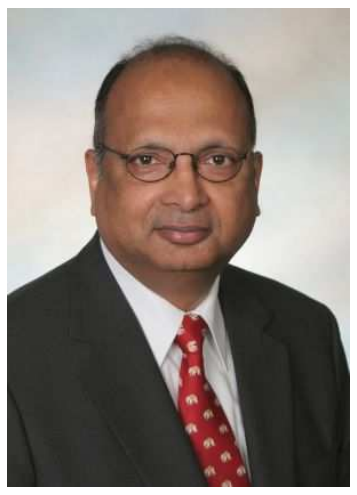

**Arogyaswami Paulraj** (SM'85-F'91) was educated at the Naval Engineering College and received the Ph.D. degree from the Indian Institute of Technology, New Delhi, India in 1973.

Currently, he is a Professor in the Department of Electrical Engineering, Stanford University, Stanford, CA, where he supervises the Smart Antennas Research Group working on applications of space-time techniques for wireless communications. His nonacademic positions have included Head of the Sonar Division, Naval Oceanographic Laboratory, Cochin, India; Director of the Center for Artificial Intelligence and Robotics, Bangalore, India; Director of the Center for Development of Advanced Computing, India; Chief Scientist of Bharat Electronics, Bangalore, India; Founder and Chief Technical Officer (CTO) of Iospan Wireless Inc.; and Co-Founder and CTO of Beceem Communications Inc. He is the author of over 300 research papers and holds over 20 patents. His research has spanned several disciplines, emphasizing estimation theory, sensor signal processing, parallel computer architectures/algorithms and space-time wireless communications. His engineering experience included development of sonar systems, massively parallel computers, and more recently broadband wireless systems.

Dr. Paulraj has won several awards for his research and engineering contributions, including the IEEE Signal Processing Society's Technical Achievement Award. He is a Member of the U.S. National Academy of Engineering, a Fellow of the IEEE, and a Member of the Indian National Academy of Engineering.